\documentclass[twocolumn,aps,pra,superscriptaddress,preprintnumbers,bibnotes]{revtex4-1}

\usepackage[T1]{fontenc}

\usepackage{natbib}
\usepackage{graphicx}
\usepackage{bm}
\usepackage{color}
\usepackage{amsmath}
\usepackage{gensymb} 

\renewcommand{\vec}[1]{\ensuremath{\bm{#1}}}

\DeclareMathOperator{\sign}{sign}

\newcommand\unit[2]{\ensuremath{#1~\mathrm{{#2}}}}

\newcommand\Ket[1]{\ensuremath{|{#1}\rangle}}

\usepackage{xspace}
\newcommand{\SLJ}[3]{{\ensuremath{{^{#1}}\mathrm{#2}_{#3}}}\xspace}
\newcommand{\TPT}{\SLJ{3}{P}{2}}
\newcommand{\TPO}{\SLJ{3}{P}{1}}
\newcommand{\TPZ}{\SLJ{3}{P}{0}}
\newcommand{\TSO}{\SLJ{3}{S}{1}}
\newcommand{\SSZ}{\SLJ{1}{S}{0}}
\newcommand{\SPO}{\SLJ{1}{P}{1}}

\newcommand{\Isotope}[2]{\ensuremath{^{{#1}}\text{{#2}}}\xspace}
\newcommand{\Sr}[1]{\Isotope{{#1}}{Sr}}

\begin{document}

\title{Fast and dense magneto-optical traps for Strontium}

\author{S. Snigirev}
\author{A. J. Park}
\author{A. Heinz}
\affiliation{
  Max-Planck-Institut f{\"u}r Quantenoptik,
  Hans-Kopfermann-Stra{\ss}e 1,
  85748 Garching, Germany}
\author{I. Bloch}
\affiliation{
  Max-Planck-Institut f{\"u}r Quantenoptik,
  Hans-Kopfermann-Stra{\ss}e 1,
  85748 Garching, Germany}
\affiliation{
  Fakult{\"a}t f{\"u}r Physik,
  Ludwig-Maximilians-Universit{\"a}t M{\"u}nchen,
  80799 M{\"u}nchen, Germany}
\author{S. Blatt}
\affiliation{
  Max-Planck-Institut f{\"u}r Quantenoptik,
  Hans-Kopfermann-Stra{\ss}e 1,
  85748 Garching, Germany}

\date{\today}

\begin{abstract}
  We improve the efficiency of sawtooth-wave-adiabatic-passage (SWAP) cooling~\cite{norcia18,muniz18,bartolotta18} for strontium atoms in three dimensions and combine it with standard narrow-line laser cooling.
  With this technique, we create strontium magneto-optical traps with $6\times 10^7$ bosonic \Sr{88}  ($1\times 10^7$ fermionic \Sr{87}) atoms at phase-space densities of $2\times 10^{-3}$ ($1.4\times 10^{-4}$).
  Our method is simple to implement and is faster and more robust than traditional cooling methods.
\end{abstract}

\maketitle

\section{Introduction}

Ultracold strontium (Sr) atoms are used in optical frequency standards~\cite{ludlow15}, in superradiant lasers~\cite{norcia18c} and atom interferometers~\cite{hu17}, for studies of molecular~\cite{barbe18,kondov18} and Rydberg~\cite{ding18} physics, to constrain the variation of fundamental constants~\cite{blatt08}, for quantum simulation~\cite{kolkowitz16,rajagopal17}, and for experiments with atom arrays~\cite{cooper18,norcia18b}.
Although continuous sources of ultracold Sr atoms are under development~\cite{bennetts17,chen18}, all these experiments  operate with a duty cycle that is limited by the sample preparation time.
This duty cycle fundamentally prevents optical clocks~\cite{itano93,ludlow15} from overcoming the standard quantum limit~\cite{braginsky92} by aliasing technical noise into the measurement results~\cite{santarelli98,nicholson15}.
High repetition rates also benefit quantum simulators with ultracold atoms~\cite{bloch12}, and are a necessary requirement to implement novel schemes such as variational quantum simulation~\cite{lanyon10,kokail18}.

With this in mind, we apply the recently-developed sawtooth-wave-adiabatic-passage~(SWAP) technique~\cite{norcia18,muniz18,bartolotta18,petersen18} to improve the performance of our narrow-line magneto-optical traps (MOTs), described in Sec.~\ref{sec:experiment}.
We model the cooling process using a moving three-level atom in the presence of the spatially varying magnetic field~\cite{muniz18} in Sec.~\ref{sec:obe}.
From this model, we conclude that the broadband cooling used in most Sr MOTs is better understood within the same adiabatic passage framework.
Nevertheless, our theoretical and experimental results show that SWAP cooling in a MOT is necessarily more robust and efficient (Sec.~\ref{sec:swap-mot}).
In free space, SWAP cooling can exploit stimulated emission to cool faster than the limit imposed by the \unit{21}{\mu s} natural lifetime of the cooling transition~\cite{bartolotta18}.
However, we find theoretically and experimentally that the spatially varying magnetic field of the MOT, in combination with polarization selection rules, prevents us from exploiting stimulated emission.
Although SWAP cooling does not make use of stimulated emission, it provides compelling benefits over traditional broadband laser cooling in a Sr MOT:
it strongly improves the sample preparation time, efficiency, and robustness for both bosonic \Sr{88} and fermionic \Sr{87} isotopes.

We build on our improved understanding of the cooling process in Sec.~\ref{sec:sf-mot} and combine three different cooling stages to create 3-$\mu$K-cold samples of bosonic \Sr{88} atoms from an initial, magnetically trapped, cloud at \unit{1}{mK} within \unit{50}{ms}.
These results are enabled by a novel SWAP MOT stage where only one axis is exposed to laser light at a time, but the illuminated axis is changed every \unit{45}{\mu s}.
This technique avoids unwanted stimulated processes between different axes and speeds up the SWAP cooling before the final  cooling stage.
Finally, we adapt our cooling method to MOTs of fermionic \Sr{87} and demonstrate the same benefits.

\begin{figure}
  \centering
  \includegraphics{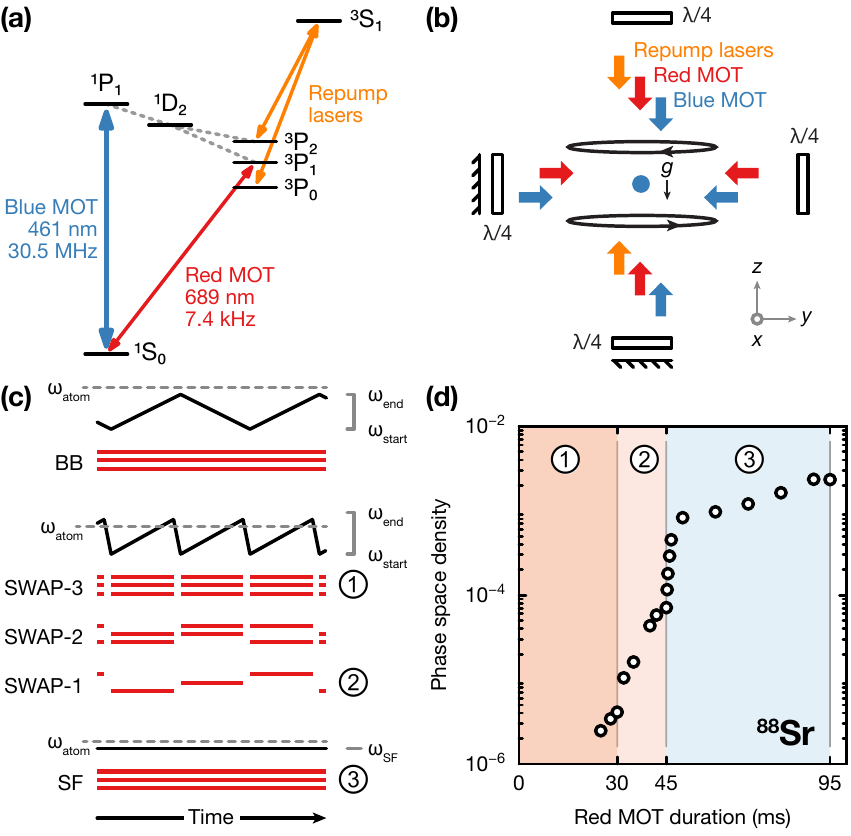}
  \caption{(color online). (a) Strontium energy level diagram and transitions used in the experiment. We use a magneto-optical trap (``Blue MOT'') on the \SSZ{}-\SPO{} transition to load a magnetic trap for the \TPT{} state. After repumping, we switch to a MOT on the \SSZ{}-\TPO{} transition (``Red MOT''). (b) Both MOTs use retroreflected beams and the repump lasers propagate along the direction of gravity $g$. (c) Detuning (black solid lines) and illumination (red rectangles) sequences for the three axes used in different cooling stages of the red MOT, as explained in the main text. (d) Combining sequences (1), (2), and (3) leads to high phase-space-density \Sr{88} samples on time scales below \unit{100}{ms}.}
  \label{fig:intro}
\end{figure}

\section{Experiment}
\label{sec:experiment}

We load strontium atoms into a magnetic trap~\cite{mukaiyama03,stellmer13} for the \TPT{} state from a Zeeman-slowed atomic beam source.
The atoms are transferred from the magneto-optical trap (``Blue MOT'') on the \SSZ{}-\SPO{} transition, with a natural linewidth $\Gamma_\mathrm{blue} = 2\pi\times\unit{30.5}{MHz}$ [see Fig.~\ref{fig:intro}(a)], to the magnetic trap.
The magnetic trap stores a dilute atomic gas at the Doppler temperature $T_D = \hbar\Gamma_\mathrm{blue}/(2 k_B) = \unit{0.7}{mK}$.
Here, $2\pi\hbar=h$ is Planck's constant and $k_B$ is Boltzmann's constant.
For the blue MOT, we use three retroreflected laser beams at \unit{460.86}{nm} with powers of (\unit{6}{mW}, \unit{6}{mW}, \unit{4}{mW}) along the ($X$, $Y$, $Z$) axes and $1/e^2$-waists of \unit{6}{mm}, as sketched in Fig.~\ref{fig:intro}(b).
A pair of anti-Helmholtz coils provides the magnetic quadrupole field $B(\rho, z) = B' \sqrt{\rho^2/4 + z^2}$
for the MOT, with a gradient $B' = \unit{63.7}{G/cm}$ ($B'/2$), with respect to the axial (transverse) coordinate $z$ ($\rho = \sqrt{x^2 + y^2}$).
These conditions lead to a trapped atom cloud in the linear potential $U(\rho, z) = g(\TPT) m(\TPT) \mu_B B(\rho, z)$, and an exponentially decaying density profile.
Here, $g(\TPT) = 3/2$ is the magnetic $g$-factor of the \TPT{} state, $m(\TPT)$ is the magnetic quantum number, and $\mu_B$ is the Bohr magneton.
The density profile for the bosonic isotope \Sr{88} thus depends on the relative occupation of the magnetic sublevels $|m| = 1$ and $2$.
The hyperfine structure due to the large nuclear spin ($I=9/2$) in the fermionic isotope \Sr{87} complicates predictions of the density profile further.
Compared to \Sr{88}, the five hyperfine states have different and much smaller $g$-factors, which lead to a more extended and less tightly trapped atomic cloud.

The atom number in the magnetic trap saturates when the gain by loading from the atomic beam balances the loss due to collisions with the atomic beam.
For our system, we find a corresponding 1/$e$ magnetic trap lifetime of \unit{24}{s} (\unit{16}{s}) for bosonic \Sr{88} (fermionic \Sr{87}) at an oven temperature of 600$\,^\circ\mathrm{C}$.
After \unit{3}{s} of loading, we apply a \unit{10}{ms} pulse of repumping laser light to the sample.
For this purpose, we use two lasers that operate on the \TPT{}-\TSO{} and \TPZ{}-\TSO{} transitions at \unit{707}{nm} and \unit{679}{nm}, respectively [see Fig.~\ref{fig:intro}(a)].
The repump pulse transfers atoms to the \TPO{} state, from which they decay with a lifetime of $\tau=\unit{21.28(3)}{\mu s}$~\cite{nicholson15} back to the \SSZ{} ground state.
For the laser intensities and magnetic fields used here, the \SSZ{} state population is refilled with a $1/e$-time of \unit{1.3(1)}{ms}.
In the spinless electronic ground state, the atoms experience almost no magnetic force and start to expand freely.

To further cool the atoms to the $\mu$K regime, they need to be captured in a secondary narrow-line magneto-optical trap (``red MOT'') operating on the \SSZ{}-\TPO{} transition at $\lambda \simeq \unit{689.4}{nm}$ with linewidth $\Gamma =
1/\tau = 2\pi\times\unit{7.48(1)}{kHz}$.
The large discrepancy between red and blue transition linewidths makes it necessary to significantly broaden the linewidth of the red MOT laser to prevent atom loss:
The Doppler-broadened linewidth $\Delta\omega_D = 2\pi\times \sqrt{4\hbar\Gamma_\text{blue}\ln{2}/ (m\lambda^2)} \simeq2\pi\times\unit{0.9}{MHz}$ is $\sim$120 times larger than $\Gamma$.
Furthermore, spatially confining atoms in a magneto-optical trap for the \SSZ{}-\TPO{} transition requires a magnetic quadrupole field with typical axial gradients $B'$ of a few G/cm~\cite{mukaiyama03,loftus04,chaneliere08,stellmer13}.
This order-of-magnitude reduction in magnetic field compared to the blue MOT has to be achieved on timescales comparable to the \SSZ{} refilling time to prevent atoms from escaping due to their per-axis atomic root-mean-square velocity $\sim$\unit{0.25}{mm/ms}.
For this reason, we switch the field gradient diabatically and we find a typical Zeeman shift of several MHz on the red MOT transition.

The traditional strategy to overcome such large Doppler and Zeeman shifts is to frequency-modulate the red MOT laser at a modulation frequency $f_\mathrm{mod}$ over a period $t_\mathrm{sweep} = 1/f_\mathrm{mod}$.
The resulting laser spectrum is a comb of frequencies spaced by $f_\mathrm{mod}$, and care has to be taken to find a balance between modulation speed and power-broadened linewidth.
Traditionally, the resulting cooling process has been explained in terms of Doppler cooling with a modified laser spectrum.
However, as we will show below, even the traditional approach is more usefully described in terms of adiabatic rapid passage processes, because optimal sweep times are comparable to the atomic lifetime~$\tau$~\cite{mukaiyama03,loftus04,chaneliere08,stellmer13}.

Specifically, we investigated the frequency modulation and illumination sequences sketched in Fig.~\ref{fig:intro}(c).
In the first strategy, we use broadband-modulated laser cooling (BB), similar to traditional frequency-comb Doppler cooling.
Here, the laser frequency is scanned in a triangle ramp between $\omega_\mathrm{start}$ and $\omega_\mathrm{end}$, such that the laser is always red-detuned from the atomic resonance at $\omega_\mathrm{atom}$.
We use three retroreflected laser beams that are always turned on, as indicated by the continuous illumination sequence below the frequency scan in Fig.~\ref{fig:intro}(c).
The red MOT lasers have $1/e^2$-waists of \unit{3}{mm} and we use powers of up to \unit{8}{mW} per beam.
All measurements in this paper use red light derived from a tapered amplifier, seeded with a diode laser that is itself stabilized to a high-finesse reference cavity.

As an alternative to the BB strategy, we investigate the  sawtooth-wave adiabatic passage (SWAP) cooling technique~\cite{norcia18,muniz18,bartolotta18}.
In this method, the laser frequency is ramped in a sawtooth-shaped ramp, as shown in the center panel of Fig.~\ref{fig:intro}(c).
In contrast to BB, the laser is swept across the free-space atomic resonance to $\omega_\mathrm{end}$ and is rapidly reset to $\omega_\mathrm{start}$ on a timescale that is fundamentally limited by the acoustic wave transfer time in the acousto-optical modulators that we use.
To avoid another sweep across the resonance during this reset, we turn off the radio-frequency power in the acousto-optic modulators at $\omega_\mathrm{end}$.
In combination with technical limitations in the timing system, the frequency reset results in dark time of $\sim$\unit{5}{\mu s} after each sweep.

We investigate three SWAP strategies, labelled SWAP-3, SWAP-2, and SWAP-1, respectively, corresponding to the number of bright axes during each frequency sweep.
Here, SWAP-3 is the only previously studied strategy in the context of magneto-optical trapping~\cite{muniz18}.
After a period of frequency-modulated laser cooling according to these strategies, we apply a period of red-detuned single-frequency Doppler cooling to the atoms, indicated in the bottom panel of Fig.~\ref{fig:intro}(c) as strategy SF.
We find that a combination of the strategies SWAP-3, SWAP-1, and SF results in the highest phase-space-density samples on the shortest time scales.
We show typical data for the bosonic \Sr{88} isotope in Fig.~\ref{fig:intro}(d), where we gain three orders of magnitude in phase space density in less than \unit{100}{ms}.
Note that most of the cooling happens within the first \unit{50}{ms} of our combined sequence, which uses the SWAP-3, SWAP-1, and SF strategies consecutively.

We find that reducing the SWAP cooling to a uniaxial process with SWAP-1 results in higher speed and final phase space density, but requires pre-cooling the atoms with the highest Doppler and Zeeman shifts to avoid losing them when an axis is dark.
In the following Section, we develop a simple model to explain both BB and SWAP strategies within a common framework.

\section{Cooling Model}
\label{sec:obe}

\begin{figure*}
  \centering
  \includegraphics{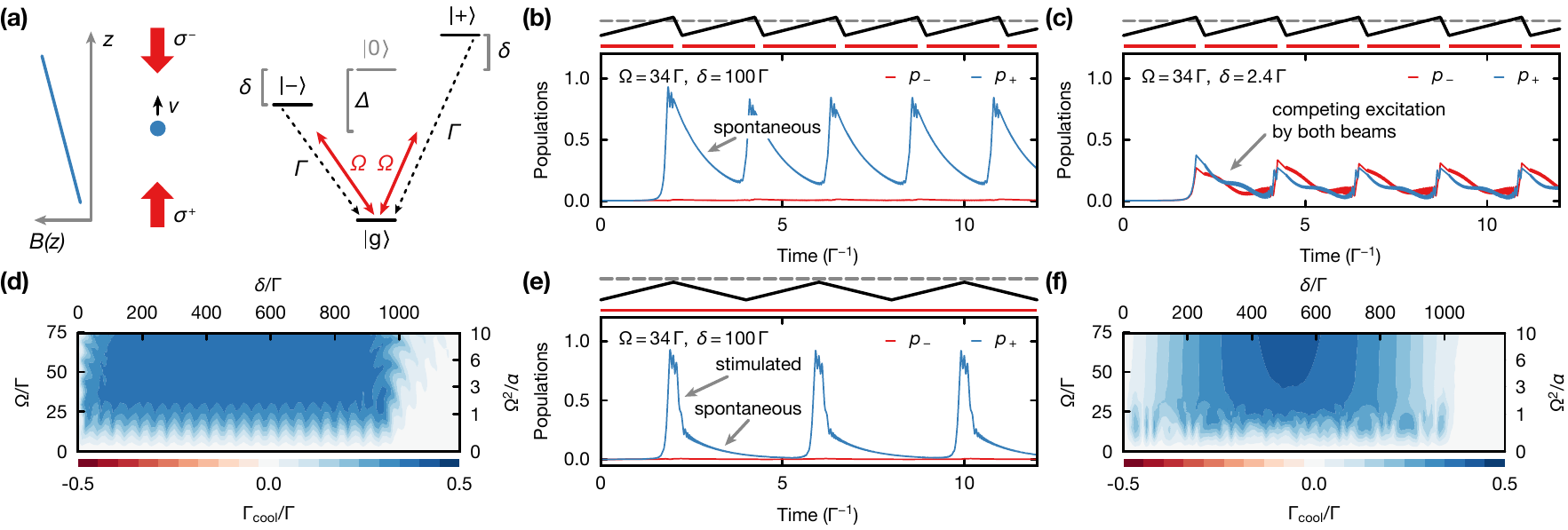}
  \caption{(color online). (a) One-dimensional laser cooling configuration in the presence of a magnetic field gradient. We use a reduced three-level system in a V configuration to model cooling on the \Sr{88} \SSZ{}-\TPO{} transition. (b) Typical population dynamics in the high-velocity (or $|\delta| \gg \Omega/\sqrt{2}$) regime. We use $t_\mathrm{sweep} = 2\tau$ and $\Delta_\mathrm{sweep} = 1000\,\Gamma$ for all results shown in this Figure. (c) Typical population dynamics in the low-velocity (or $|\delta| \ll \Omega/\sqrt{2}$) regime where cooling stops. (d) In the adiabatic regime, where $\Omega^2/\alpha > 2/\pi$, the cooling rate $\Gamma_\mathrm{cool}$ is remarkably insensitive to the level splitting $\delta$ and Rabi frequency $\Omega$. (e, f) Traditional broadband frequency-modulated cooling can be understood within the same framework. For atoms at small $|\delta|$, the downward sweep causes stimulated emission by the \emph{same} beam that caused the excitation on the upward sweep. This process partially cancels the desired momentum transfer, reduces the cooling rate, and causes a stronger parameter dependence in the low-$|\delta|$ regime for BB compared to SWAP.}
  \label{fig:threelevel}
\end{figure*}

We are interested in the average behavior of a thermal sample of three-level \Sr{88} atoms interacting with a train of frequency-swept laser pulses on the \SSZ{}-\TPO{} transition in the presence of a quadrupole magnetic field.
Based on our experimental results, we will argue later that the population dynamics for \Sr{87} with its ten nuclear magnetic states can be understood in a similar framework.
To model the atom-light interaction, we use a simplified model first introduced for this purpose in Ref.~\cite{muniz18}.

Specifically, we include the non-degenerate \SSZ{} ground state \Ket{g} and the two stretched magnetic sublevels $\Ket{\pm}$ of the \TPO{} state (V-type level scheme) in the optical Bloch equations for an atom moving along one dimension, say $Z$.
Two laser beams with equal intensities and opposite circular polarizations propagate with wave vectors $\pm k \hat{\vec{z}}$, where $k=2\pi/\lambda$, as sketched in Fig.~\ref{fig:threelevel}(a).

We treat the atomic position $z$ and velocity $v$ classically and thus can combine the Doppler and Zeeman shifts of \Ket{\pm} into a single parameter $\delta = kv + g(\TPO) m(\TPO) \mu_B B' z/\hbar$ that describes the energy splitting between the states \Ket{\pm} corresponding to the magnetic quantum numbers $m(\TPO) = \pm 1$.
The $J=0 \to J=1$ transition under consideration leads to equal Clebsch-Gordan factors of $1/\sqrt{3}$ for all possible transitions.
Although we use retroreflected laser beams, which produce a standing wave with rotating linear polarization at each position, $|\delta| > 0$ locally selects the resonant transition and
the cooling process terminates as soon as $|\delta|$ locally becomes small compared to the power-broadened linewidth.
For a magnetic quadrupole field, the atom is thus cooled to a drift velocity pointing towards the magnetic field zero.

The above considerations result in an equal Rabi frequency $\Omega \equiv \Gamma/\sqrt{3} \sqrt{s_0/2}$ for each beam.  Here, $s_0 = I_\mathrm{pk} / I_\mathrm{sat}$ is the saturation parameter in terms of the saturation intensity $I_\mathrm{sat} = \pi h c / (3 \lambda^3 \tau)$ and the Gaussian laser beams' peak intensity $I_\mathrm{pk} = 2 P / (\pi w_0^2)$, with beam power $P$ and $1/e^2$-waist $w_0$, respectively.
We also allow for the lasers to be switched off by letting $\Omega(t)$ vary with time.
The laser frequency for each beam is scanned simultaneously as $\Delta(t) \equiv \Delta_0 + f(t)$, starting at a fixed initial detuning $\Delta_0 \equiv \omega_\mathrm{start} - \omega_\mathrm{atom}$ and continuing with a periodic frequency ramp $f(t)$.

Under these assumptions, we find the time-dependent Hamiltonian
\begin{equation}
  \label{eq:2}
  H(t)/\hbar =
  \begin{pmatrix}
    \Delta(t) + \delta & 0 & \Omega(t)/2 \\
    0 & \Delta(t) - \delta & \Omega(t)/2 \\
    \Omega^*(t)/2 & \Omega^*(t)/2 & 0 \\
  \end{pmatrix},
\end{equation}
and the optical Bloch equations for the density matrix $\rho$
\begin{equation}
  \label{eq:3}
  \dot\rho = -i [H(t)/\hbar, \rho] + \mathcal{L}\rho.
\end{equation}
We model the effects of spontaneous emission on the elements of the density matrix by the Liouvillian
\begin{equation}
  \label{eq:4}
  \mathcal{L}\rho = -\Gamma
  \begin{pmatrix}
    \rho_{11} & \rho_{12} & \rho_{13}/2 \\
    \rho_{21} & \rho_{22} & \rho_{23}/2 \\
    \rho_{31}/2 & \rho_{32}/2 & -\rho_{11} - \rho_{22} \\
  \end{pmatrix}.
\end{equation}

This model is useful to describe the loading and initial cooling of the red MOT because the atomic velocity and position do not change significantly on the timescale of the cycle time $t_\mathrm{cycle} \equiv t_\mathrm{sweep} + t_\mathrm{dark}$, which in all cases of interest is on the order of the atomic lifetime $\tau$.
This condition places the initial stage of frequency-swept laser cooling in the red MOT in an interesting regime.
We work neither in the adiabatic rapid passage regime,
where $t_\mathrm{cycle} \ll \tau$, nor fully in the steady state with respect to atomic decay, where $t_\mathrm{cycle} \gg \tau$.
For this reason, adiabatic approximations of the Bloch equations produce misleading results and we have to rely on numerical solutions to explain our experimental results.
For instance, we show the population dynamics of a typical pulse train for a representative sweep (dead) time of $t_\mathrm{sweep} = 2 \tau$ ($t_\mathrm{dead} = 0.238 \tau$) in the high velocity regime in Fig.~\ref{fig:threelevel}(b).
Here, an atom at detuning $\delta = 100~\Gamma$ is exposed to a train of laser pulses whose frequency is swept over $\Delta_\mathrm{sweep} = 1000~\Gamma$, ending at  $\omega_\mathrm{end} - \omega_\mathrm{atom} = +13.3~\Gamma$, with a Rabi frequency of $\Omega = 34~\Gamma$.
Because of the large splitting between the excited states, the pulse train efficiently excites only the \Ket{+} state.
After the first excitation, spontaneous emission reinitializes the atom to a ground state fraction depending on $t_\mathrm{cycle}/\tau$.
We find that the population dynamics reliably settle to a periodic pattern for all parameter ranges in this work after a few cycles.

Even though we have to use numerics, we can identify useful analytic expressions for some of the parameters, such as the condition for adiabatic passage, if spontaneous emission is neglected.
Assuming that $\Omega$ is constant and that the detuning is ramped across the resonance with constant frequency slope $\alpha \equiv \dot{f} = \Delta_\mathrm{sweep}/t_\mathrm{sweep}$, the Landau-Zener probability for adiabatic passage to the excited state~\cite{bartolotta18}
\begin{equation}
  \label{eq:5}
  p_\mathrm{LZ} = 1 - \exp{\left(-\frac{\pi}{2}\frac{\Omega^2}{\alpha}\right)},
\end{equation}
is only determined by the adiabaticity parameter $\Omega^2/\alpha$.
Note again that this result requires $t_\mathrm{cycle} \ll \tau$, but that it will be useful to benchmark our experimental and numerical results.
In particular, the excited state population never reaches $p_\mathrm{LZ}$, because it decays during the whole excitation process.

We can also see from Eqn.~\ref{eq:2} that if $|\delta| \ll \Omega/\sqrt{2}$, because the velocity and the Zeeman splitting are small or compensate each other, we have a competition between adiabatic passage from the ground state to either of the excited states \Ket{\pm}.
If there is no imbalance between the transition probability to \Ket{\pm}, the cooling efficiency vanishes, because the atom absorbs a photon from each of the counterpropagating beams.
Typical population dynamics for $\delta = 2.4~\Gamma$ are shown in Fig.~\ref{fig:threelevel}(c).

The transition from cooling to heating leads to a balance where one finds the same steady-state temperature $k_B T_\mathrm{ss} = \hbar\Omega/2$ as for Doppler cooling as long as one cannot take advantage of stimulated processes where the atom is stimulated back to the ground state by the \emph{other} beam~\cite{bartolotta18}.
In contrast to optical molasses~\cite{norcia18} it is not possible to realize this situation in a magneto-optical trap~\cite{muniz18}, because opposite circular polarizations are used in combination with a magnetic-field gradient to create localization.
In a situation where one can separate atomic localization from the excitation process, such as in a magic-wavelength optical dipole trap, SWAP cooling could be much more effective by exploiting stimulated emission in the regime of $t_\mathrm{cycle} \ll \tau$ as originally envisioned~\cite{norcia18,bartolotta18}.

To describe the efficiency of the cooling process, we introduce the laser cooling rate
\begin{equation}
  \label{eq:6}
  \Gamma_\mathrm{cool} \equiv \Gamma \sign(\delta) \langle p_+ - p_- \rangle_\mathrm{cycle}
\end{equation}
as the difference between the scattering rates due to the cycle-averaged probabilities of exciting the corresponding states $p_+ \equiv \rho_{11}$ and $p_- \equiv \rho_{22}$, respectively.
Because the SWAP cooling process is based on adiabatic passage, this cooling rate is remarkably insensitive to laser frequency or intensity drifts, as shown in Fig.~\ref{fig:threelevel}(d).

Interestingly, we can understand the broadband-modulated laser cooling (BB), traditionally used in narrow-line magneto-optical traps for Sr~\cite{mukaiyama03,loftus04,stellmer13} within the same framework:
In Fig.~\ref{fig:threelevel}(e), we show population dynamics for a pulse train where the laser frequency is ramped in a triangle pattern with the same slope (Rabi frequency) $\alpha$ ($\Omega$) as in panels (b) and (c), such that the adiabaticity parameter remains the same.
The laser is never turned off ($t_\mathrm{dead} = 0$) and the detuning ramp still spans $\Delta_\mathrm{sweep}=1000~\Gamma$, but ends to the red of the resonance at $\omega_\mathrm{end} - \omega_\mathrm{atom} = -13.3~\Gamma$.
We immediately see the disadvantage of this BB strategy compared to the SWAP-1 strategy, in that $p_+$ is not allowed to decay spontaneously, but is stimulated back to the ground state on the down-slope of the ramp by the \emph{same} beam that excited it.
This stimulated process produces a momentum kick opposite to the initial excitation and reduces the amount of spontaneous scattering, and thus $\Gamma_\mathrm{cool}$.

As shown in Fig.~\ref{fig:threelevel}(f), the BB strategy works well when the time between adiabatic transfers on the up- and down-slope of the frequency ramp is long enough for a significant fraction of $p_+$ to decay, because adiabatic passage is insensitive to the direction of the frequency sweep across the resonance.
However, the cooling efficiency is strongly reduced for low-$|\delta|$ atoms compared to SWAP-1.
Some of this efficiency can be recovered by modulating the laser frequency in a sinusoidal fashion (reduced $\alpha$ at small $|\delta|$) as traditionally done~\cite{loftus04,chaneliere08}, but SWAP is more efficient.

In conclusion, we find that the adiabatic passage picture provides a better framework to understand both traditional BB and SWAP cooling strategies.
In addition, the model predicts that, compared to BB, SWAP (1) optimizes the excitation process for low-velocity atoms at low Zeeman shifts, (2) makes the cooling process more homogeneous across the whole thermal sample loaded from the magnetic trap, and (3) is more robust with respect to intensity fluctuations.

In the subsequent section, we will show experimental results that support this conclusion and discuss secondary experimental conditions that influence the choice of cooling strategy.

\section{SWAP MOT}
\label{sec:swap-mot}

\begin{figure}
  \centering
  \includegraphics{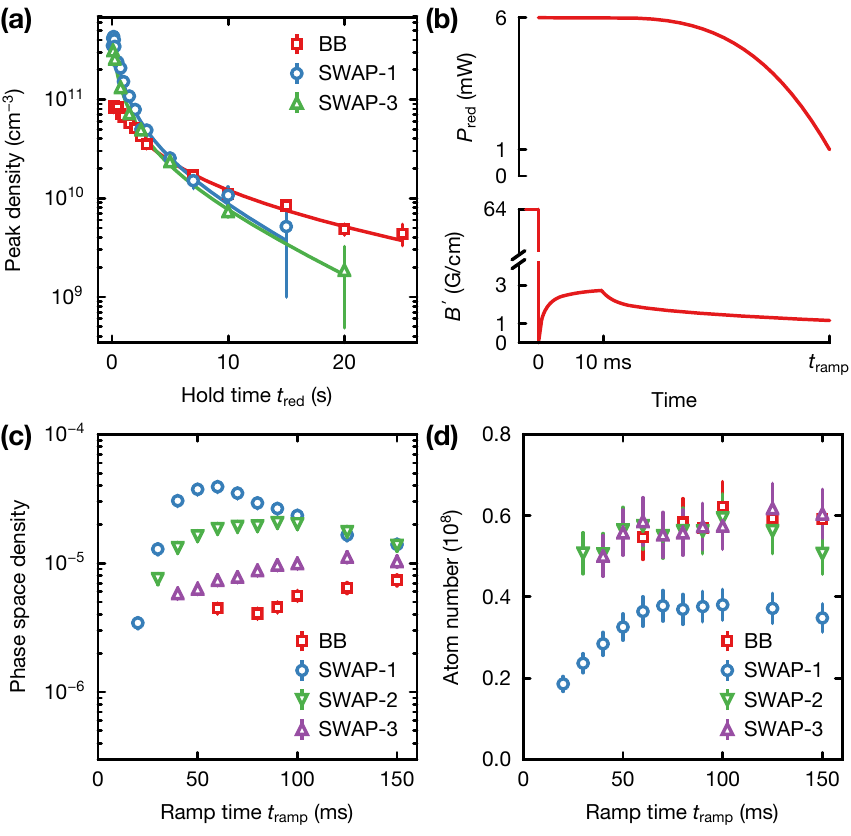}
  \caption{(color online). (a) The peak density decreases as a function of hold time when the intensity and magnetic field gradient are held constant. At short times, light-assisted collisions at high densities lead to loss for all cooling strategies. (b) Optimized ramp of the light intensity and magnetic field gradient used to measure phase-space-densities (c) and atom numbers (d) versus ramp time $t_\mathrm{ramp}$ for all modulation strategies.}
  \label{fig:mots88}
\end{figure}

To study the differences between the cooling strategies sketched in Fig.~\ref{fig:intro}(c), we start by applying the corresponding pulse train for a time $t_\mathrm{red}$ to the atomic sample while keeping the magnetic field gradient and the laser intensity constant.
At the end of $t_\mathrm{red}$, we turn off the magnetic field gradient as well as the laser beams, and either image the atoms in situ, or allow the atoms to fall for \unit{15}{ms} before imaging.
We take two absorption images simultaneously by exposing the atomic cloud for \unit{50}{\mu s} to two separate probe beams propagating along $Y$ and $Z$, respectively.
Using standard methods~\cite{blatt11}, we extract the temperature, atom number, and in-trap phase-space-density of the atomic cloud.
The error bars for these quantities combine a 10\% shot-to-shot atom number fluctuation with the statistical fit error derived by rescaling each image fit to $\chi^2 = 1$.

The results for varying cooling times $t_\mathrm{red}$ are shown in Fig.~\ref{fig:mots88}(a).
Here, all strategies use a common red laser power of \unit{2}{mW} per axis and a sweep range of $\sim$\unit{11}{MHz}.
The SWAP strategies end at a (blue) detuning of \unit{100}{kHz}, while the BB strategy ends at a (red) detuning of $-\unit{100}{kHz}$.
We use a sweep time $t_\mathrm{sweep} = \unit{40}{\mu s}$ (\unit{80}{\mu s}) for SWAP (BB).

Compared to the SWAP strategies, BB exhibits a lower initial density, but a slower decay at long times.
We determine both $1/e$ lifetime $\tau_\mathrm{MOT}$ and two-body-loss rate coefficient $K_2$ for all strategies by fitting the solution of $\dot{n} = - n/\tau_\mathrm{MOT} - K_2 n^2$ to the density data in Fig.~\ref{fig:mots88}(a).
We find that the red-detuned BB strategy leads to a MOT with $\tau_\mathrm{MOT} = \unit{25 \pm 10 }{s}$, comparable to the lifetime of atoms in the magnetic trap.
We thus attribute this one-body loss to collisions with the atomic beam.
The SWAP-1 and SWAP-3 strategies lead to a reduced $\tau_\mathrm{MOT} = \unit{7(2)}{s}$ and \unit{8(2)}{s}, respectively.
In addition, all strategies show non-exponential loss at short times, due to light-assisted scattering on the repulsive $V_{1u}$ asymptote~\cite{zelevinsky06}.
We find similar two-body-loss rate coefficients  $K_2\simeq\unit{5(1)\times 10^{-12}}{cm^3/s}$ for all strategies at this laser power.

Previous attempts at optimizing the broadband stage of the cooling procedure made a choice between quickly cooling only the coldest atoms for atomic clocks~\cite{loftus04} and slowly cooling almost all atoms for quantum gas experiments~\cite{stellmer13}.
With the SWAP technique, we aimed to combine the advantages of both methods and varied the parameters of each strategy to obtain the coldest samples in the shortest times.
As we see in Fig.~\ref{fig:mots88}(a), the SWAP-1 strategy can condense hot atoms on fast timescales and thus reaches its steady-state temperature quickly.
This steady-state temperature is proportional to the laser intensity, and the shape of the atomic cloud is determined by the magnetic field gradient~\cite{loftus04}.
We thus ramp both magnetic field gradient and laser intensity with the empirically optimized polynomial shapes shown in Fig.~\ref{fig:mots88}(b) while the atoms are cooled.
Under these conditions, we find that the strategies produce samples with dramatically different phase-space densities as a function of total ramp time.
In Fig.~\ref{fig:mots88}(c), we see that all strategies have an optimal associated time:
If we ramp too quickly, the phase-space density remains low.
If we ramp too slowly, we start to lose phase-space density due to light-assisted collisions between the coldest atoms.
We also see that SWAP-1 produces the highest phase-space densities while BB performs the worst.
The SWAP-1 strategy achieves this goal despite losing 40\% of the atoms, as shown in Fig.~\ref{fig:mots88}(d).
This loss is not present in the other strategies, and we attribute this loss to hot atoms that escape from the cooling region while the corresponding axes are not illuminated.

Based on these results, we decided to combine the high capture efficiency of SWAP-3 with the fast and efficient cooling of SWAP-1.
We use the same laser power and magnetic field ramps as in Fig.~\ref{fig:mots88}(b), but switch from SWAP-3 to SWAP-1 at a time $t_\mathrm{switch} < t_\mathrm{ramp}$.
We optimized $t_\mathrm{switch}$ and the SWAP cooling parameters of this combined sequence in detail for both bosonic \Sr{88} and fermionic \Sr{87} isotopes, and found that its performance is limited by the initial capture fraction of SWAP-3 from the magnetic trap.

\begin{figure}
  \centering
  \includegraphics{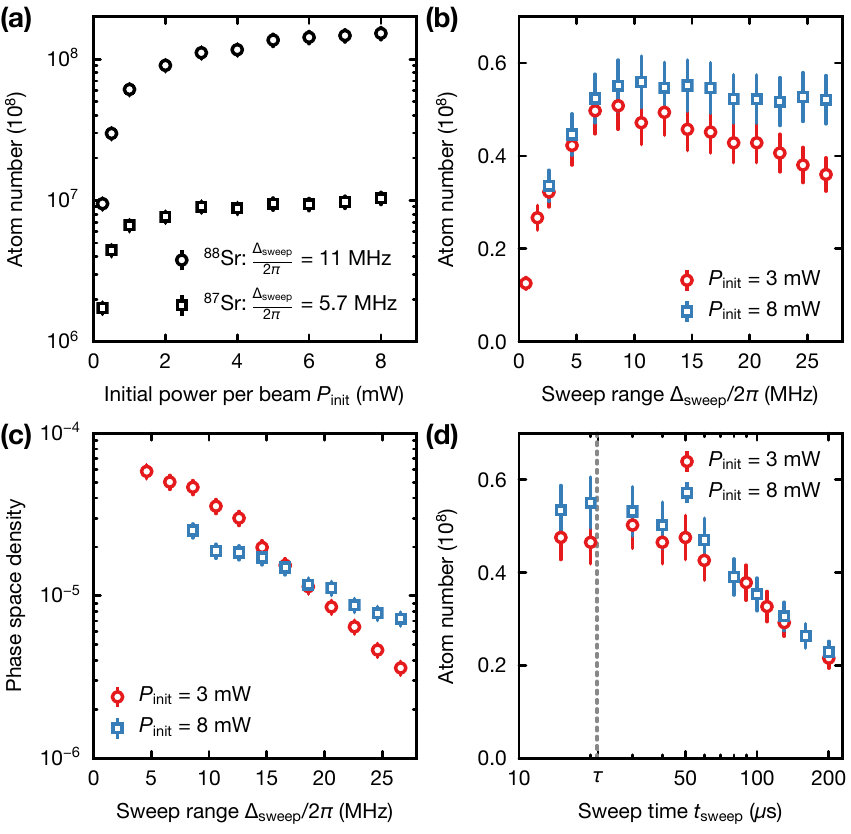}
  \caption{(color online). (a) The atom number in the SWAP MOT for an \unit{8.5}{MHz} sweep range saturates as a function of the initial laser power per beam $P_\mathrm{init}$ for \Sr{88} and \Sr{87}. (b) For \Sr{88}, the atom number saturates for high powers per beam (blue squares) as a function of sweep range, but decreases linearly for large sweep range at low powers (red circles). (c) At the same time, the phase-space density decreases exponentially. (d) Longer sweep times preclude capturing the fastest atoms from the magnetic trap.  Sweep times shorter than the natural lifetime $\tau$ do not increase the capture fraction further.}
  \label{fig:params}
\end{figure}

In Fig.~\ref{fig:params}(a), we show the atom number at $t_\mathrm{ramp}=\unit{45}{ms}$ (\unit{150}{ms}) for \Sr{88} (\Sr{87}) versus the initial power per beam $P_\mathrm{init}$.
We trap $1.5\times10^{8}$ ($1.0\times10^7$) \Sr{88} (\Sr{87}) atoms for $P_\mathrm{init} = \unit{8}{mW}$.
The data suggests that we reach the adiabatic passage regime for relatively low initial powers.
We find that a sweep range of $\Delta_\mathrm{sweep} = 2\pi\times\unit{11}{MHz}$ ($2\pi\times\unit{5.7}{MHz}$) for \Sr{88} (\Sr{87}) produces a comparable power-dependence for both isotopes.
The ratio between sweep ranges is consistent with similar cooling conditions requiring similar adiabaticity parameters, and the lower average scattering rate for the $F=9/2 \to F'=11/2$ transition in \Sr{87} compared to the $J=0 \to J'=1$ transition in \Sr{88}.
The final number of \Sr{87} atoms is $\sim$80\% of the value suggested by the relative natural abundance of \Sr{87} and \Sr{88} (7.00\%/82.58\%).
We attribute this discrepancy to the more extended atomic density profile in the magnetic trap (see Sec.~\ref{sec:experiment}).
If sufficient optical power is available, increasing the beam sizes could lead to an improved capture fraction.
For a given beam size, the capture fraction of the SWAP MOT seems to be proportional to the adiabaticity parameter if we take into account that \Sr{87} scatters less cooling light than \Sr{88}.

In the remainder of Fig.~\ref{fig:params}, we explore the SWAP cooling parameters for two representative initial powers: (1) a ``low'' power per beam of \unit{3}{mW} that is available from a typical diode laser at \unit{689}{nm}, and (2) a ``high'' power per beam of \unit{8}{mW} that requires multiple diode lasers or a tapered amplifier.
For brevity, we only show data for \Sr{88}, because we find  equivalent results for \Sr{87} with the caveat of a reduced scattering rate that requires a reduced sweep range for the same power.

\begin{figure}
  \centering
  \includegraphics{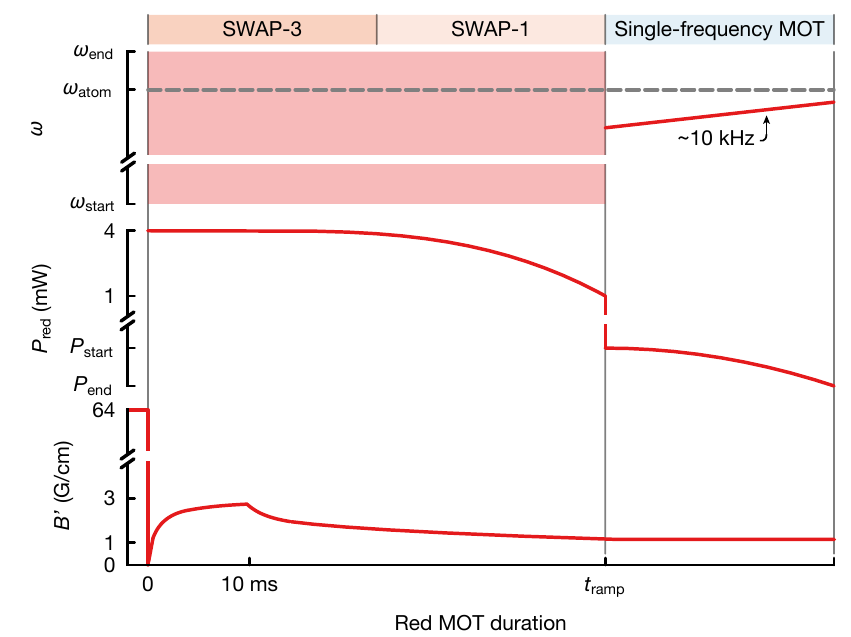}
  \caption{Optimized experimental sequence used for our both \Sr{88} and \Sr{87} magneto-optical traps. The cooling technique in a given interval is indicated at the top (SWAP-3, SWAP-1 or single-frequency MOT). Top, middle, and bottom graphs show the red MOT beam frequency spectrum, power, and gradient traces versus the red MOT duration, respectively.}
  \label{fig:ramp_shape}
\end{figure}

\begin{figure*}
  \centering
  \includegraphics{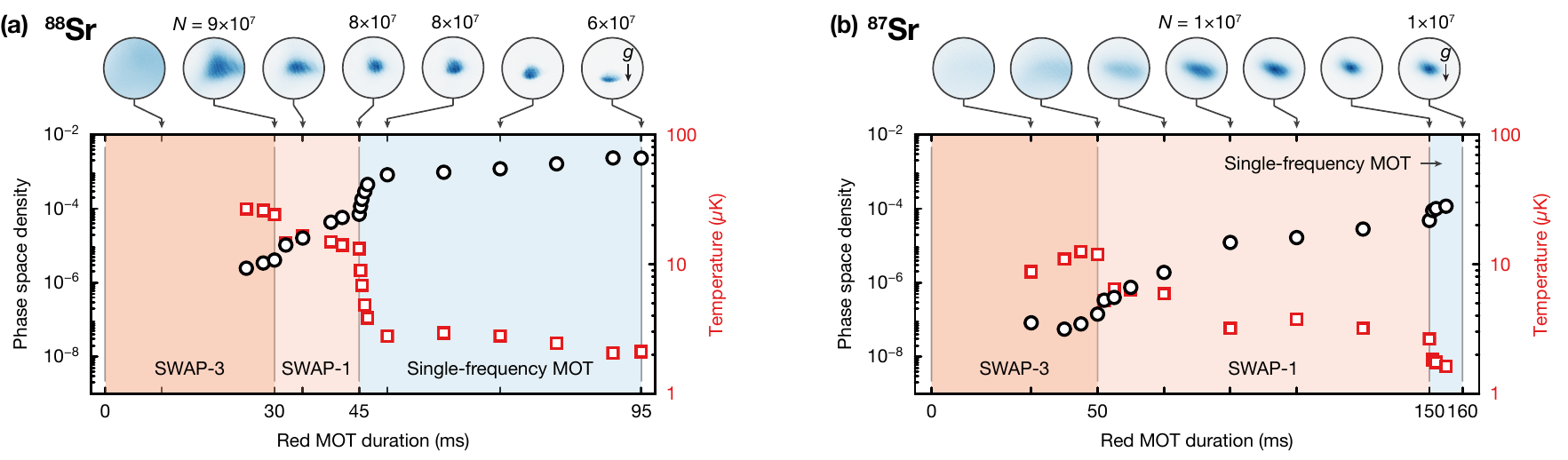}
  \caption{(a) Measured temperatures (red squares) and phase-space densities (black circles) versus red MOT duration for \Sr{88}. The label on the bottom (SWAP-3, SWAP-1 or single-frequency MOT) specifies the active cooling strategy.  In-situ images taken at different red MOT times are shown on the top with atom number ($N$) and the direction of gravity ($g$). (b) We find comparable results for \Sr{87} when taking the reduced scattering rate into account.}
  \label{fig:sr88_single_frequency}
\end{figure*}

When we vary the sweep range $\Delta_\mathrm{sweep}$ by varying $\omega_\mathrm{start}$, we find the data shown in Fig.~\ref{fig:params}(b).
For high power, the atom number first increases and then saturates because an increased sweep range can address atoms at higher Zeeman shifts.
For low power, the atom number peaks, but then slowly decreases with the linear decrease in adiabaticity parameter.
Even though the atom number shows a similar behavior in the low- and high-power limits, the phase space density decreases exponentially with increased sweep range, as shown in Fig.~\ref{fig:params}(c).
This behavior is consistent with an exponential decrease in the cooling rate due to the reduced adiabatic transfer efficiency $\propto p_\mathrm{LZ}$.
We show in Fig.~\ref{fig:params}(d) that the sweep time influences the number of atoms dramatically as well: the cooling rate is too small to capture the fastest atoms for increased sweep times.
Finally, we find that reducing the sweep time below the natural lifetime does not improve the number of captured atoms in the SWAP MOT, consistent with the predictions of the optical Bloch equations in Sec.~\ref{sec:obe}.

\section{Single-frequency MOT}
\label{sec:sf-mot}

As a last step in our cooling protocol, we use traditional narrow-line laser cooling at a single frequency to reach final temperatures of \unit{1-2}{\mu K}.
We start the red MOT with the optimized SWAP combination sequence discussed in Sec.~\ref{sec:swap-mot}.
The laser frequency is scanned from $\omega_\mathrm{start}-\omega_\mathrm{atom} = -2\pi\times\unit{8.5}{MHz} \ (-\unit{4.2}{MHz})$ to $\omega_\mathrm{end}-\omega_\mathrm{atom} = 2\pi\times\unit{0.1}{MHz}$ for \Sr{88} (\Sr{87}) as shown in the upper panel of Fig.~\ref{fig:ramp_shape}.
At the same time, the laser power and magnetic field gradient are slowly ramped with the polynomial shapes shown in Fig.~\ref{fig:ramp_shape}.
After reaching the steady state of the combined technique, we switch to the single-frequency MOT at $t_{\textrm{ramp}}$ to further cool the sample.
To switch to the SF strategy, we select cooling parameters that would leave the cloud shape and temperature unchanged.
Thus, we turn off the frequency scan, set the laser frequency to a $-\unit{80}{kHz}$ ($-\unit{10}{kHz}$) red detuning, and quickly lower the beam power from \unit{1}{mW} to $P_{\mathrm{start}} = \unit{35}{\mu W}$ (\unit{20}{\mu W}) for \Sr{88} (\Sr{87}).
Finally, we ramp the beam power once again with a polynomial shape to $P_{\mathrm{end}} = \unit{1}{\mu W}$ (\unit{0.5}{\mu W}) for \Sr{88} (\Sr{87}) to reduce the steady-state temperature of the cooling process.
To ensure fast cooling during the single-frequency MOT, we minimize the atomic movement along gravity caused by the change in the detuning and gradient~\cite{loftus04}.
We thus limit the detuning ramp amplitude to only $\sim$\unit{10}{kHz} and keep the gradient constant.

The series of in-situ absorption images of \Sr{88} in Fig.~\ref{fig:sr88_single_frequency}(a) illustrates the cooling process.
The SWAP-3 strategy allows us to capture about $9\times10^7$ atoms, but the cloud remains large and dilute.
As soon as we switch to SWAP-1, the atomic cloud shrinks visibly.
In the single-frequency MOT, the atoms sag along the direction of gravity while cooling to a few $\mu$K, which is a characteristic behavior of the bosonic narrow-line MOT~\cite{loftus04}.
We cool to \unit{3}{\mu K} after \unit{5}{ms} of the single-frequency MOT without losing atoms.
The phase-space density at this point is $8\times 10^{-4}$, a factor of 400 larger than for the case of a red MOT time of \unit{25}{ms} (prior to this, we cannot get reliable estimates due to irregular in-situ shapes).
The phase-space density increases further over the final \unit{45}{ms} of single-frequency cooling and reaches $2\times 10^{-3}$ with a final temperature of \unit{2}{\mu K} at the expense of losing 25\% of the atoms.
Note that this final cooling step in the single-frequency MOT takes the same amount of time as all of the initial cooling procedure, pointing towards a mechanism that competes with the cooling process while the \Sr{88} atoms sag to the lower edge of the MOT.
This atom loss is likely due to a combination of light-assisted collisions and radiation trapping~\cite{katori99,yang15}.

We apply the same protocol to \Sr{87}, but increase its cooling efficiency by adding red stirring laser beams ~\cite{mukaiyama03}, which copropagate with the red MOT beams.
The in-situ images in Fig.~\ref{fig:sr88_single_frequency}(b) show the cooling progress for \Sr{87}.
Unlike \Sr{88} with its vanishingly small scattering length, the \Sr{87} sample does not sag under gravity.
Instead, it thermalizes by interparticle collisions~\cite{mukaiyama03,stellmer13}.
We observe larger and more dilute initial atomic clouds of \Sr{87} than of \Sr{88} during SWAP-3, because of the reduced cooling rate discussed in the previous Section.
For the same reason, it takes longer to condense the \Sr{87} cloud to the steady state in the subsequent SWAP-1 cooling stage.
In total, we find that we need to operate the SWAP MOT about three times longer for \Sr{87} than for \Sr{88}.
We reach a temperature of \unit{3}{\mu K} and a phase-space density of $5\times 10^{-5}$ at the end of SWAP-1.
The subsequent \unit{10}{ms} of single-frequency MOT cools the atoms further, reaching a final temperature of \unit{1.4}{\mu K} and a phase-space density of $1.4\times 10^{-4}$ without atom loss.

\section{Conclusion}
\label{sec:conclusion}

We have demonstrated that sawtooth-wave adiabatic-passage (SWAP) can be used to create high phase-space-density samples of bosonic (fermionic) \Sr{88} (\Sr{87}) atoms within \unit{50}{ms} (\unit{160}{ms}).
Extending our method by a final dark-spot MOT stage~\cite{stellmer13b} might result in even lower final temperatures and higher phase-space densities.
Our results suggest that the narrow-line, single-frequency cooling stage produces most of its effect on timescales of \unit{10}{ms} before it becomes limited by density-dependent effects~\cite{katori99,yang15}.
Our method is simple to implement and provides a useful improvement over the traditional broadband cooling stage in terms of speed and robustness.
In combination with high-flux atomic sources~\cite{nagel05,mickelson05,yang15}, our method can be used to improve the duty cycle of atomic clocks, and the repetition rate of precision experiments and quantum simulations.
Our method can also improve narrow-line magneto-optical traps for other two-electron atoms, lanthanides, or molecules.

We thank N.~Jan\v{s}a, R.~Gonz\'{a}lez~Escudero, S.~Wissenberg, F.~Finger, R.~Haindl, E.~Staub, A.~Mayer, and K.~F\"{o}rster for technical contributions to the construction of the experiment, and N.~\v{S}anti\'c for critical reading of the manuscript. This work was supported by funding from the UQUAM ERC Synergy Grant. A.~J.~P. was supported by a fellowship from the Natural Sciences and Engineering Research Council of Canada (NSERC).


\begin{thebibliography}{35}%
\makeatletter
\providecommand \@ifxundefined [1]{%
 \@ifx{#1\undefined}
}%
\providecommand \@ifnum [1]{%
 \ifnum #1\expandafter \@firstoftwo
 \else \expandafter \@secondoftwo
 \fi
}%
\providecommand \@ifx [1]{%
 \ifx #1\expandafter \@firstoftwo
 \else \expandafter \@secondoftwo
 \fi
}%
\providecommand \natexlab [1]{#1}%
\providecommand \enquote  [1]{``#1''}%
\providecommand \bibnamefont  [1]{#1}%
\providecommand \bibfnamefont [1]{#1}%
\providecommand \citenamefont [1]{#1}%
\providecommand \href@noop [0]{\@secondoftwo}%
\providecommand \href [0]{\begingroup \@sanitize@url \@href}%
\providecommand \@href[1]{\@@startlink{#1}\@@href}%
\providecommand \@@href[1]{\endgroup#1\@@endlink}%
\providecommand \@sanitize@url [0]{\catcode `\\12\catcode `\$12\catcode
  `\&12\catcode `\#12\catcode `\^12\catcode `\_12\catcode `\%12\relax}%
\providecommand \@@startlink[1]{}%
\providecommand \@@endlink[0]{}%
\providecommand \url  [0]{\begingroup\@sanitize@url \@url }%
\providecommand \@url [1]{\endgroup\@href {#1}{\urlprefix }}%
\providecommand \urlprefix  [0]{URL }%
\providecommand \Eprint [0]{\href }%
\providecommand \doibase [0]{http://dx.doi.org/}%
\providecommand \selectlanguage [0]{\@gobble}%
\providecommand \bibinfo  [0]{\@secondoftwo}%
\providecommand \bibfield  [0]{\@secondoftwo}%
\providecommand \translation [1]{[#1]}%
\providecommand \BibitemOpen [0]{}%
\providecommand \bibitemStop [0]{}%
\providecommand \bibitemNoStop [0]{.\EOS\space}%
\providecommand \EOS [0]{\spacefactor3000\relax}%
\providecommand \BibitemShut  [1]{\csname bibitem#1\endcsname}%
\let\auto@bib@innerbib\@empty
\bibitem [{\citenamefont {Norcia}\ \emph
  {et~al.}(2018{\natexlab{a}})\citenamefont {Norcia}, \citenamefont {Cline},
  \citenamefont {Bartolotta}, \citenamefont {Holland},\ and\ \citenamefont
  {Thompson}}]{norcia18}%
  \BibitemOpen
  \bibfield  {author} {\bibinfo {author} {\bibfnamefont {M.~A.}\ \bibnamefont
  {Norcia}}, \bibinfo {author} {\bibfnamefont {J.~R.~K.}\ \bibnamefont
  {Cline}}, \bibinfo {author} {\bibfnamefont {J.~P.}\ \bibnamefont
  {Bartolotta}}, \bibinfo {author} {\bibfnamefont {M.~J.}\ \bibnamefont
  {Holland}}, \ and\ \bibinfo {author} {\bibfnamefont {J.~K.}\ \bibnamefont
  {Thompson}},\ }\href {\doibase 10.1088/1367-2630/aaa950} {\bibfield
  {journal} {\bibinfo  {journal} {N. J. Phys.}\ }\textbf {\bibinfo {volume}
  {20}},\ \bibinfo {pages} {023021} (\bibinfo {year}
  {2018}{\natexlab{a}})}\BibitemShut {NoStop}%
\bibitem [{\citenamefont {Muniz}\ \emph {et~al.}()\citenamefont {Muniz},
  \citenamefont {Norcia}, \citenamefont {Cline},\ and\ \citenamefont
  {Thompson}}]{muniz18}%
  \BibitemOpen
  \bibfield  {author} {\bibinfo {author} {\bibfnamefont {J.~A.}\ \bibnamefont
  {Muniz}}, \bibinfo {author} {\bibfnamefont {M.~A.}\ \bibnamefont {Norcia}},
  \bibinfo {author} {\bibfnamefont {J.~R.~K.}\ \bibnamefont {Cline}}, \ and\
  \bibinfo {author} {\bibfnamefont {J.~K.}\ \bibnamefont {Thompson}},\
  }\href@noop {} {\ }\Eprint {http://arxiv.org/abs/1806.00838v1}
  {arXiv:1806.00838v1} \BibitemShut {NoStop}%
\bibitem [{\citenamefont {Bartolotta}\ \emph {et~al.}(2018)\citenamefont
  {Bartolotta}, \citenamefont {Norcia}, \citenamefont {Cline}, \citenamefont
  {Thompson},\ and\ \citenamefont {Holland}}]{bartolotta18}%
  \BibitemOpen
  \bibfield  {author} {\bibinfo {author} {\bibfnamefont {J.~P.}\ \bibnamefont
  {Bartolotta}}, \bibinfo {author} {\bibfnamefont {M.~A.}\ \bibnamefont
  {Norcia}}, \bibinfo {author} {\bibfnamefont {J.~R.~K.}\ \bibnamefont
  {Cline}}, \bibinfo {author} {\bibfnamefont {J.~K.}\ \bibnamefont {Thompson}},
  \ and\ \bibinfo {author} {\bibfnamefont {M.~J.}\ \bibnamefont {Holland}},\
  }\href {\doibase 10.1103/PhysRevA.98/023404} {\bibfield  {journal} {\bibinfo
  {journal} {Phys. Rev. A}\ }\textbf {\bibinfo {volume} {98}},\ \bibinfo
  {pages} {023404} (\bibinfo {year} {2018})}\BibitemShut {NoStop}%
\bibitem [{\citenamefont {Ludlow}\ \emph {et~al.}(2015)\citenamefont {Ludlow},
  \citenamefont {Boyd}, \citenamefont {Ye}, \citenamefont {Peik},\ and\
  \citenamefont {Schmidt}}]{ludlow15}%
  \BibitemOpen
  \bibfield  {author} {\bibinfo {author} {\bibfnamefont {A.~D.}\ \bibnamefont
  {Ludlow}}, \bibinfo {author} {\bibfnamefont {M.~M.}\ \bibnamefont {Boyd}},
  \bibinfo {author} {\bibfnamefont {J.}~\bibnamefont {Ye}}, \bibinfo {author}
  {\bibfnamefont {E.}~\bibnamefont {Peik}}, \ and\ \bibinfo {author}
  {\bibfnamefont {P.}~\bibnamefont {Schmidt}},\ }\href {\doibase
  10.1103/RevModPhys.87.637} {\bibfield  {journal} {\bibinfo  {journal} {Rev.
  Mod. Phys.}\ }\textbf {\bibinfo {volume} {87}},\ \bibinfo {pages} {637}
  (\bibinfo {year} {2015})}\BibitemShut {NoStop}%
\bibitem [{\citenamefont {Norcia}\ \emph
  {et~al.}(2018{\natexlab{b}})\citenamefont {Norcia}, \citenamefont {Cline},
  \citenamefont {Muniz}, \citenamefont {Robinson}, \citenamefont {Hutson},
  \citenamefont {Goban}, \citenamefont {Marti}, \citenamefont {Ye},\ and\
  \citenamefont {Thompson}}]{norcia18c}%
  \BibitemOpen
  \bibfield  {author} {\bibinfo {author} {\bibfnamefont {M.~A.}\ \bibnamefont
  {Norcia}}, \bibinfo {author} {\bibfnamefont {J.~R.}\ \bibnamefont {Cline}},
  \bibinfo {author} {\bibfnamefont {J.~A.}\ \bibnamefont {Muniz}}, \bibinfo
  {author} {\bibfnamefont {J.~M.}\ \bibnamefont {Robinson}}, \bibinfo {author}
  {\bibfnamefont {R.~B.}\ \bibnamefont {Hutson}}, \bibinfo {author}
  {\bibfnamefont {A.}~\bibnamefont {Goban}}, \bibinfo {author} {\bibfnamefont
  {G.~E.}\ \bibnamefont {Marti}}, \bibinfo {author} {\bibfnamefont
  {J.}~\bibnamefont {Ye}}, \ and\ \bibinfo {author} {\bibfnamefont {J.~K.}\
  \bibnamefont {Thompson}},\ }\href {\doibase 10.1103/PhysRevX.8.021036}
  {\bibfield  {journal} {\bibinfo  {journal} {Phys. Rev. X}\ }\textbf {\bibinfo
  {volume} {8}},\ \bibinfo {pages} {021036} (\bibinfo {year}
  {2018}{\natexlab{b}})}\BibitemShut {NoStop}%
\bibitem [{\citenamefont {Hu}\ \emph {et~al.}(2017)\citenamefont {Hu},
  \citenamefont {Poli}, \citenamefont {Salvi},\ and\ \citenamefont
  {Tino}}]{hu17}%
  \BibitemOpen
  \bibfield  {author} {\bibinfo {author} {\bibfnamefont {L.}~\bibnamefont
  {Hu}}, \bibinfo {author} {\bibfnamefont {N.}~\bibnamefont {Poli}}, \bibinfo
  {author} {\bibfnamefont {L.}~\bibnamefont {Salvi}}, \ and\ \bibinfo {author}
  {\bibfnamefont {G.~M.}\ \bibnamefont {Tino}},\ }\href {\doibase
  10.1103/PhysRevLett.119.263601} {\bibfield  {journal} {\bibinfo  {journal}
  {Phys. Rev. Lett.}\ }\textbf {\bibinfo {volume} {119}},\ \bibinfo {pages}
  {263601} (\bibinfo {year} {2017})}\BibitemShut {NoStop}%
\bibitem [{\citenamefont {Barb{\'e}}\ \emph {et~al.}(2018)\citenamefont
  {Barb{\'e}}, \citenamefont {Ciamei}, \citenamefont {Pasquiou}, \citenamefont
  {Reichs{\"o}llner}, \citenamefont {Schreck}, \citenamefont {Żuchowski},\
  and\ \citenamefont {Hutson}}]{barbe18}%
  \BibitemOpen
  \bibfield  {author} {\bibinfo {author} {\bibfnamefont {V.}~\bibnamefont
  {Barb{\'e}}}, \bibinfo {author} {\bibfnamefont {A.}~\bibnamefont {Ciamei}},
  \bibinfo {author} {\bibfnamefont {B.}~\bibnamefont {Pasquiou}}, \bibinfo
  {author} {\bibfnamefont {L.}~\bibnamefont {Reichs{\"o}llner}}, \bibinfo
  {author} {\bibfnamefont {F.}~\bibnamefont {Schreck}}, \bibinfo {author}
  {\bibfnamefont {P.~S.}\ \bibnamefont {Żuchowski}}, \ and\ \bibinfo {author}
  {\bibfnamefont {J.~M.}\ \bibnamefont {Hutson}},\ }\href {\doibase
  10.1038/s41567-018-0169-x} {\bibfield  {journal} {\bibinfo  {journal} {Nat.
  Phys.}\ }\textbf {\bibinfo {volume} {14}},\ \bibinfo {pages} {881} (\bibinfo
  {year} {2018})}\BibitemShut {NoStop}%
\bibitem [{\citenamefont {Kondov}\ \emph {et~al.}(2018)\citenamefont {Kondov},
  \citenamefont {Lee}, \citenamefont {McDonald}, \citenamefont {McGuyer},
  \citenamefont {Majewska}, \citenamefont {Moszynski},\ and\ \citenamefont
  {Zelevinsky}}]{kondov18}%
  \BibitemOpen
  \bibfield  {author} {\bibinfo {author} {\bibfnamefont {S.}~\bibnamefont
  {Kondov}}, \bibinfo {author} {\bibfnamefont {C.-H.}\ \bibnamefont {Lee}},
  \bibinfo {author} {\bibfnamefont {M.}~\bibnamefont {McDonald}}, \bibinfo
  {author} {\bibfnamefont {B.}~\bibnamefont {McGuyer}}, \bibinfo {author}
  {\bibfnamefont {I.}~\bibnamefont {Majewska}}, \bibinfo {author}
  {\bibfnamefont {R.}~\bibnamefont {Moszynski}}, \ and\ \bibinfo {author}
  {\bibfnamefont {T.}~\bibnamefont {Zelevinsky}},\ }\href {\doibase
  10.1103/PhysRevLett.121.143401} {\bibfield  {journal} {\bibinfo  {journal}
  {Phys. Rev. Lett.}\ }\textbf {\bibinfo {volume} {121}},\ \bibinfo {pages}
  {143401} (\bibinfo {year} {2018})}\BibitemShut {NoStop}%
\bibitem [{\citenamefont {Ding}\ \emph {et~al.}(2018)\citenamefont {Ding},
  \citenamefont {Whalen}, \citenamefont {Kanungo}, \citenamefont {Killian},
  \citenamefont {Dunning}, \citenamefont {Yoshida},\ and\ \citenamefont
  {Burgd{\"o}rfer}}]{ding18}%
  \BibitemOpen
  \bibfield  {author} {\bibinfo {author} {\bibfnamefont {R.}~\bibnamefont
  {Ding}}, \bibinfo {author} {\bibfnamefont {J.~D.}\ \bibnamefont {Whalen}},
  \bibinfo {author} {\bibfnamefont {S.~K.}\ \bibnamefont {Kanungo}}, \bibinfo
  {author} {\bibfnamefont {T.~C.}\ \bibnamefont {Killian}}, \bibinfo {author}
  {\bibfnamefont {F.~B.}\ \bibnamefont {Dunning}}, \bibinfo {author}
  {\bibfnamefont {S.}~\bibnamefont {Yoshida}}, \ and\ \bibinfo {author}
  {\bibfnamefont {J.}~\bibnamefont {Burgd{\"o}rfer}},\ }\href {\doibase
  10.1103/PhysRevA.98.042505} {\bibfield  {journal} {\bibinfo  {journal} {Phys.
  Rev. A}\ }\textbf {\bibinfo {volume} {98}},\ \bibinfo {pages} {042505}
  (\bibinfo {year} {2018})}\BibitemShut {NoStop}%
\bibitem [{\citenamefont {Blatt}\ \emph {et~al.}(2008)\citenamefont {Blatt},
  \citenamefont {Ludlow}, \citenamefont {Campbell}, \citenamefont {Thomsen},
  \citenamefont {Zelevinsky}, \citenamefont {Boyd}, \citenamefont {Ye},
  \citenamefont {Baillard}, \citenamefont {Fouch{\'e}}, \citenamefont {Targat},
  \citenamefont {Brusch}, \citenamefont {Lemonde}, \citenamefont {Takamoto},
  \citenamefont {Hong}, \citenamefont {Katori},\ and\ \citenamefont
  {Flambaum}}]{blatt08}%
  \BibitemOpen
  \bibfield  {author} {\bibinfo {author} {\bibfnamefont {S.}~\bibnamefont
  {Blatt}}, \bibinfo {author} {\bibfnamefont {A.}~\bibnamefont {Ludlow}},
  \bibinfo {author} {\bibfnamefont {G.}~\bibnamefont {Campbell}}, \bibinfo
  {author} {\bibfnamefont {J.}~\bibnamefont {Thomsen}}, \bibinfo {author}
  {\bibfnamefont {T.}~\bibnamefont {Zelevinsky}}, \bibinfo {author}
  {\bibfnamefont {M.}~\bibnamefont {Boyd}}, \bibinfo {author} {\bibfnamefont
  {J.}~\bibnamefont {Ye}}, \bibinfo {author} {\bibfnamefont {X.}~\bibnamefont
  {Baillard}}, \bibinfo {author} {\bibfnamefont {M.}~\bibnamefont
  {Fouch{\'e}}}, \bibinfo {author} {\bibfnamefont {R.~L.}\ \bibnamefont
  {Targat}}, \bibinfo {author} {\bibfnamefont {A.}~\bibnamefont {Brusch}},
  \bibinfo {author} {\bibfnamefont {P.}~\bibnamefont {Lemonde}}, \bibinfo
  {author} {\bibfnamefont {M.}~\bibnamefont {Takamoto}}, \bibinfo {author}
  {\bibfnamefont {F.-L.}\ \bibnamefont {Hong}}, \bibinfo {author}
  {\bibfnamefont {H.}~\bibnamefont {Katori}}, \ and\ \bibinfo {author}
  {\bibfnamefont {V.}~\bibnamefont {Flambaum}},\ }\href {\doibase
  10.1103/PhysRevLett.100.140801} {\bibfield  {journal} {\bibinfo  {journal}
  {Phys. Rev. Lett.}\ }\textbf {\bibinfo {volume} {100}},\ \bibinfo {pages}
  {140801} (\bibinfo {year} {2008})}\BibitemShut {NoStop}%
\bibitem [{\citenamefont {Kolkowitz}\ \emph {et~al.}(2016)\citenamefont
  {Kolkowitz}, \citenamefont {Bromley}, \citenamefont {Bothwell}, \citenamefont
  {Wall}, \citenamefont {Marti}, \citenamefont {Koller}, \citenamefont {Zhang},
  \citenamefont {Rey},\ and\ \citenamefont {Ye}}]{kolkowitz16}%
  \BibitemOpen
  \bibfield  {author} {\bibinfo {author} {\bibfnamefont {S.}~\bibnamefont
  {Kolkowitz}}, \bibinfo {author} {\bibfnamefont {S.~L.}\ \bibnamefont
  {Bromley}}, \bibinfo {author} {\bibfnamefont {T.}~\bibnamefont {Bothwell}},
  \bibinfo {author} {\bibfnamefont {M.~L.}\ \bibnamefont {Wall}}, \bibinfo
  {author} {\bibfnamefont {G.~E.}\ \bibnamefont {Marti}}, \bibinfo {author}
  {\bibfnamefont {A.~P.}\ \bibnamefont {Koller}}, \bibinfo {author}
  {\bibfnamefont {X.}~\bibnamefont {Zhang}}, \bibinfo {author} {\bibfnamefont
  {A.~M.}\ \bibnamefont {Rey}}, \ and\ \bibinfo {author} {\bibfnamefont
  {J.}~\bibnamefont {Ye}},\ }\href {\doibase 10.1038/nature20811} {\bibfield
  {journal} {\bibinfo  {journal} {Nature}\ }\textbf {\bibinfo {volume} {542}},\
  \bibinfo {pages} {66} (\bibinfo {year} {2016})}\BibitemShut {NoStop}%
\bibitem [{\citenamefont {Rajagopal}\ \emph {et~al.}(2017)\citenamefont
  {Rajagopal}, \citenamefont {Fujiwara}, \citenamefont {Senaratne},
  \citenamefont {Singh}, \citenamefont {Geiger},\ and\ \citenamefont
  {Weld}}]{rajagopal17}%
  \BibitemOpen
  \bibfield  {author} {\bibinfo {author} {\bibfnamefont {S.~V.}\ \bibnamefont
  {Rajagopal}}, \bibinfo {author} {\bibfnamefont {K.~M.}\ \bibnamefont
  {Fujiwara}}, \bibinfo {author} {\bibfnamefont {R.}~\bibnamefont {Senaratne}},
  \bibinfo {author} {\bibfnamefont {K.}~\bibnamefont {Singh}}, \bibinfo
  {author} {\bibfnamefont {Z.~A.}\ \bibnamefont {Geiger}}, \ and\ \bibinfo
  {author} {\bibfnamefont {D.~M.}\ \bibnamefont {Weld}},\ }\href {\doibase
  10.1002/andp.201700008} {\bibfield  {journal} {\bibinfo  {journal} {Ann.
  Phys.}\ }\textbf {\bibinfo {volume} {529}},\ \bibinfo {pages} {1700008}
  (\bibinfo {year} {2017})}\BibitemShut {NoStop}%
\bibitem [{\citenamefont {Cooper}\ \emph {et~al.}(2018)\citenamefont {Cooper},
  \citenamefont {Covey}, \citenamefont {Madjarov}, \citenamefont {Porsev},
  \citenamefont {Safronova},\ and\ \citenamefont {Endres}}]{cooper18}%
  \BibitemOpen
  \bibfield  {author} {\bibinfo {author} {\bibfnamefont {A.}~\bibnamefont
  {Cooper}}, \bibinfo {author} {\bibfnamefont {J.~P.}\ \bibnamefont {Covey}},
  \bibinfo {author} {\bibfnamefont {I.~S.}\ \bibnamefont {Madjarov}}, \bibinfo
  {author} {\bibfnamefont {S.~G.}\ \bibnamefont {Porsev}}, \bibinfo {author}
  {\bibfnamefont {M.~S.}\ \bibnamefont {Safronova}}, \ and\ \bibinfo {author}
  {\bibfnamefont {M.}~\bibnamefont {Endres}},\ }\href {\doibase
  10.1103/PhysRevX.8.041055} {\bibfield  {journal} {\bibinfo  {journal} {Phys.
  Rev. X}\ }\textbf {\bibinfo {volume} {8}},\ \bibinfo {pages} {041055}
  (\bibinfo {year} {2018})}\BibitemShut {NoStop}%
\bibitem [{\citenamefont {Norcia}\ \emph
  {et~al.}(2018{\natexlab{c}})\citenamefont {Norcia}, \citenamefont {Young},\
  and\ \citenamefont {Kaufman}}]{norcia18b}%
  \BibitemOpen
  \bibfield  {author} {\bibinfo {author} {\bibfnamefont {M.}~\bibnamefont
  {Norcia}}, \bibinfo {author} {\bibfnamefont {A.}~\bibnamefont {Young}}, \
  and\ \bibinfo {author} {\bibfnamefont {A.}~\bibnamefont {Kaufman}},\ }\href
  {\doibase 10.1103/PhysRevX.8.041054} {\bibfield  {journal} {\bibinfo
  {journal} {Phys. Rev. X}\ }\textbf {\bibinfo {volume} {8}},\ \bibinfo {pages}
  {041054} (\bibinfo {year} {2018}{\natexlab{c}})}\BibitemShut {NoStop}%
\bibitem [{\citenamefont {Bennetts}\ \emph {et~al.}(2017)\citenamefont
  {Bennetts}, \citenamefont {Chen}, \citenamefont {Pasquiou},\ and\
  \citenamefont {Schreck}}]{bennetts17}%
  \BibitemOpen
  \bibfield  {author} {\bibinfo {author} {\bibfnamefont {S.}~\bibnamefont
  {Bennetts}}, \bibinfo {author} {\bibfnamefont {C.-C.}\ \bibnamefont {Chen}},
  \bibinfo {author} {\bibfnamefont {B.}~\bibnamefont {Pasquiou}}, \ and\
  \bibinfo {author} {\bibfnamefont {F.}~\bibnamefont {Schreck}},\ }\href
  {\doibase 10.1103/PhysRevLett.119.223202} {\bibfield  {journal} {\bibinfo
  {journal} {Phys. Rev. Lett.}\ }\textbf {\bibinfo {volume} {119}},\ \bibinfo
  {pages} {223202} (\bibinfo {year} {2017})}\BibitemShut {NoStop}%
\bibitem [{\citenamefont {Chen}\ \emph {et~al.}()\citenamefont {Chen},
  \citenamefont {Bennetts}, \citenamefont {Escudero}, \citenamefont {Schreck},\
  and\ \citenamefont {Pasquiou}}]{chen18}%
  \BibitemOpen
  \bibfield  {author} {\bibinfo {author} {\bibfnamefont {C.-C.}\ \bibnamefont
  {Chen}}, \bibinfo {author} {\bibfnamefont {S.}~\bibnamefont {Bennetts}},
  \bibinfo {author} {\bibfnamefont {R.~G.}\ \bibnamefont {Escudero}}, \bibinfo
  {author} {\bibfnamefont {F.}~\bibnamefont {Schreck}}, \ and\ \bibinfo
  {author} {\bibfnamefont {B.}~\bibnamefont {Pasquiou}},\ }\href@noop {} {\
  }\Eprint {http://arxiv.org/abs/1810.07157v2} {arXiv:1810.07157v2}
  \BibitemShut {NoStop}%
\bibitem [{\citenamefont {Itano}\ \emph {et~al.}(1993)\citenamefont {Itano},
  \citenamefont {Bergquist}, \citenamefont {Bollinger}, \citenamefont
  {Gilligan}, \citenamefont {Heinzen}, \citenamefont {Moore}, \citenamefont
  {Raizen},\ and\ \citenamefont {Wineland}}]{itano93}%
  \BibitemOpen
  \bibfield  {author} {\bibinfo {author} {\bibfnamefont {W.~M.}\ \bibnamefont
  {Itano}}, \bibinfo {author} {\bibfnamefont {J.~C.}\ \bibnamefont
  {Bergquist}}, \bibinfo {author} {\bibfnamefont {J.~J.}\ \bibnamefont
  {Bollinger}}, \bibinfo {author} {\bibfnamefont {J.~M.}\ \bibnamefont
  {Gilligan}}, \bibinfo {author} {\bibfnamefont {D.~J.}\ \bibnamefont
  {Heinzen}}, \bibinfo {author} {\bibfnamefont {F.~L.}\ \bibnamefont {Moore}},
  \bibinfo {author} {\bibfnamefont {M.~G.}\ \bibnamefont {Raizen}}, \ and\
  \bibinfo {author} {\bibfnamefont {D.~J.}\ \bibnamefont {Wineland}},\ }\href
  {\doibase 10.1103/PhysRevA.47.3554} {\bibfield  {journal} {\bibinfo
  {journal} {Phys. Rev. A}\ }\textbf {\bibinfo {volume} {47}},\ \bibinfo
  {pages} {3554} (\bibinfo {year} {1993})}\BibitemShut {NoStop}%
\bibitem [{\citenamefont {Braginsky}\ and\ \citenamefont
  {Khalili}(1992)}]{braginsky92}%
  \BibitemOpen
  \bibfield  {author} {\bibinfo {author} {\bibfnamefont {V.~B.}\ \bibnamefont
  {Braginsky}}\ and\ \bibinfo {author} {\bibfnamefont {F.~Y.}\ \bibnamefont
  {Khalili}},\ }\href@noop {} {\emph {\bibinfo {title} {Quantum
  measurement}}},\ edited by\ \bibinfo {editor} {\bibfnamefont {K.~S.}\
  \bibnamefont {Thorne}}\ (\bibinfo  {publisher} {Cambridge University Press,
  Cambridge},\ \bibinfo {year} {1992})\BibitemShut {NoStop}%
\bibitem [{\citenamefont {Santarelli}\ \emph {et~al.}(1998)\citenamefont
  {Santarelli}, \citenamefont {Audoin}, \citenamefont {Makdissi}, \citenamefont
  {Laurent}, \citenamefont {Dick},\ and\ \citenamefont
  {Clairon}}]{santarelli98}%
  \BibitemOpen
  \bibfield  {author} {\bibinfo {author} {\bibfnamefont {G.}~\bibnamefont
  {Santarelli}}, \bibinfo {author} {\bibfnamefont {C.}~\bibnamefont {Audoin}},
  \bibinfo {author} {\bibfnamefont {A.}~\bibnamefont {Makdissi}}, \bibinfo
  {author} {\bibfnamefont {P.}~\bibnamefont {Laurent}}, \bibinfo {author}
  {\bibfnamefont {G.}~\bibnamefont {Dick}}, \ and\ \bibinfo {author}
  {\bibfnamefont {A.}~\bibnamefont {Clairon}},\ }\href {\doibase
  10.1109/58.710548} {\bibfield  {journal} {\bibinfo  {journal} {IEEE Trans.
  Ultrason. Ferroelectr. Freq. Control}\ }\textbf {\bibinfo {volume} {45}},\
  \bibinfo {pages} {887} (\bibinfo {year} {1998})}\BibitemShut {NoStop}%
\bibitem [{\citenamefont {Nicholson}\ \emph {et~al.}(2015)\citenamefont
  {Nicholson}, \citenamefont {Campbell}, \citenamefont {Hutson}, \citenamefont
  {Marti}, \citenamefont {Bloom}, \citenamefont {McNally}, \citenamefont
  {Zhang}, \citenamefont {Barrett}, \citenamefont {Safronova}, \citenamefont
  {Strouse}, \citenamefont {Tew},\ and\ \citenamefont {Ye}}]{nicholson15}%
  \BibitemOpen
  \bibfield  {author} {\bibinfo {author} {\bibfnamefont {T.}~\bibnamefont
  {Nicholson}}, \bibinfo {author} {\bibfnamefont {S.}~\bibnamefont {Campbell}},
  \bibinfo {author} {\bibfnamefont {R.}~\bibnamefont {Hutson}}, \bibinfo
  {author} {\bibfnamefont {G.}~\bibnamefont {Marti}}, \bibinfo {author}
  {\bibfnamefont {B.}~\bibnamefont {Bloom}}, \bibinfo {author} {\bibfnamefont
  {R.}~\bibnamefont {McNally}}, \bibinfo {author} {\bibfnamefont
  {W.}~\bibnamefont {Zhang}}, \bibinfo {author} {\bibfnamefont
  {M.}~\bibnamefont {Barrett}}, \bibinfo {author} {\bibfnamefont
  {M.}~\bibnamefont {Safronova}}, \bibinfo {author} {\bibfnamefont
  {G.}~\bibnamefont {Strouse}}, \bibinfo {author} {\bibfnamefont
  {W.}~\bibnamefont {Tew}}, \ and\ \bibinfo {author} {\bibfnamefont
  {J.}~\bibnamefont {Ye}},\ }\href {\doibase 10.1038/ncomms7896} {\bibfield
  {journal} {\bibinfo  {journal} {Nat. Commun.}\ }\textbf {\bibinfo {volume}
  {6}},\ \bibinfo {pages} {6896} (\bibinfo {year} {2015})}\BibitemShut
  {NoStop}%
\bibitem [{\citenamefont {Bloch}\ \emph {et~al.}(2012)\citenamefont {Bloch},
  \citenamefont {Dalibard},\ and\ \citenamefont {Nascimb{\`e}ne}}]{bloch12}%
  \BibitemOpen
  \bibfield  {author} {\bibinfo {author} {\bibfnamefont {I.}~\bibnamefont
  {Bloch}}, \bibinfo {author} {\bibfnamefont {J.}~\bibnamefont {Dalibard}}, \
  and\ \bibinfo {author} {\bibfnamefont {S.}~\bibnamefont {Nascimb{\`e}ne}},\
  }\href {\doibase 10.1038/nphys2259} {\bibfield  {journal} {\bibinfo
  {journal} {Nature Physics}\ }\textbf {\bibinfo {volume} {8}},\ \bibinfo
  {pages} {267} (\bibinfo {year} {2012})}\BibitemShut {NoStop}%
\bibitem [{\citenamefont {Lanyon}\ \emph {et~al.}(2010)\citenamefont {Lanyon},
  \citenamefont {Whitfield}, \citenamefont {Gillett}, \citenamefont {Goggin},
  \citenamefont {Almeida}, \citenamefont {Kassal}, \citenamefont {Biamonte},
  \citenamefont {Mohseni}, \citenamefont {Powell}, \citenamefont {Barbieri},
  \citenamefont {Aspuru-Guzik},\ and\ \citenamefont {White}}]{lanyon10}%
  \BibitemOpen
  \bibfield  {author} {\bibinfo {author} {\bibfnamefont {B.~P.}\ \bibnamefont
  {Lanyon}}, \bibinfo {author} {\bibfnamefont {J.~D.}\ \bibnamefont
  {Whitfield}}, \bibinfo {author} {\bibfnamefont {G.~G.}\ \bibnamefont
  {Gillett}}, \bibinfo {author} {\bibfnamefont {M.~E.}\ \bibnamefont {Goggin}},
  \bibinfo {author} {\bibfnamefont {M.~P.}\ \bibnamefont {Almeida}}, \bibinfo
  {author} {\bibfnamefont {I.}~\bibnamefont {Kassal}}, \bibinfo {author}
  {\bibfnamefont {J.~D.}\ \bibnamefont {Biamonte}}, \bibinfo {author}
  {\bibfnamefont {M.}~\bibnamefont {Mohseni}}, \bibinfo {author} {\bibfnamefont
  {B.~J.}\ \bibnamefont {Powell}}, \bibinfo {author} {\bibfnamefont
  {M.}~\bibnamefont {Barbieri}}, \bibinfo {author} {\bibfnamefont
  {A.}~\bibnamefont {Aspuru-Guzik}}, \ and\ \bibinfo {author} {\bibfnamefont
  {A.~G.}\ \bibnamefont {White}},\ }\href {\doibase 10.1038/nchem.483}
  {\bibfield  {journal} {\bibinfo  {journal} {Nat. Chem.}\ }\textbf {\bibinfo
  {volume} {2}},\ \bibinfo {pages} {106} (\bibinfo {year} {2010})}\BibitemShut
  {NoStop}%
\bibitem [{\citenamefont {Kokail}\ \emph {et~al.}()\citenamefont {Kokail},
  \citenamefont {Maier}, \citenamefont {van Bijnen}, \citenamefont {Brydges},
  \citenamefont {Joshi}, \citenamefont {Jurcevic}, \citenamefont {Muschik},
  \citenamefont {Silvi}, \citenamefont {Blatt}, \citenamefont {Roos},\ and\
  \citenamefont {Zoller}}]{kokail18}%
  \BibitemOpen
  \bibfield  {author} {\bibinfo {author} {\bibfnamefont {C.}~\bibnamefont
  {Kokail}}, \bibinfo {author} {\bibfnamefont {C.}~\bibnamefont {Maier}},
  \bibinfo {author} {\bibfnamefont {R.}~\bibnamefont {van Bijnen}}, \bibinfo
  {author} {\bibfnamefont {T.}~\bibnamefont {Brydges}}, \bibinfo {author}
  {\bibfnamefont {M.~K.}\ \bibnamefont {Joshi}}, \bibinfo {author}
  {\bibfnamefont {P.}~\bibnamefont {Jurcevic}}, \bibinfo {author}
  {\bibfnamefont {C.~A.}\ \bibnamefont {Muschik}}, \bibinfo {author}
  {\bibfnamefont {P.}~\bibnamefont {Silvi}}, \bibinfo {author} {\bibfnamefont
  {R.}~\bibnamefont {Blatt}}, \bibinfo {author} {\bibfnamefont {C.~F.}\
  \bibnamefont {Roos}}, \ and\ \bibinfo {author} {\bibfnamefont
  {P.}~\bibnamefont {Zoller}},\ }\href@noop {} {\ }\Eprint
  {http://arxiv.org/abs/1810.03421v1} {arXiv:1810.03421v1} \BibitemShut
  {NoStop}%
\bibitem [{\citenamefont {Petersen}\ \emph {et~al.}()\citenamefont {Petersen},
  \citenamefont {M{\"u}hlbauer}, \citenamefont {Bougas}, \citenamefont
  {Sharma}, \citenamefont {Budker},\ and\ \citenamefont
  {Windpassinger}}]{petersen18}%
  \BibitemOpen
  \bibfield  {author} {\bibinfo {author} {\bibfnamefont {N.}~\bibnamefont
  {Petersen}}, \bibinfo {author} {\bibfnamefont {F.}~\bibnamefont
  {M{\"u}hlbauer}}, \bibinfo {author} {\bibfnamefont {L.}~\bibnamefont
  {Bougas}}, \bibinfo {author} {\bibfnamefont {A.}~\bibnamefont {Sharma}},
  \bibinfo {author} {\bibfnamefont {D.}~\bibnamefont {Budker}}, \ and\ \bibinfo
  {author} {\bibfnamefont {P.}~\bibnamefont {Windpassinger}},\ }\href@noop {}
  {\ }\Eprint {http://arxiv.org/abs/1809.06423v1} {arXiv:1809.06423v1}
  \BibitemShut {NoStop}%
\bibitem [{\citenamefont {Mukaiyama}\ \emph {et~al.}(2003)\citenamefont
  {Mukaiyama}, \citenamefont {Katori}, \citenamefont {Ido}, \citenamefont
  {Li},\ and\ \citenamefont {Kuwata-Gonokami}}]{mukaiyama03}%
  \BibitemOpen
  \bibfield  {author} {\bibinfo {author} {\bibfnamefont {T.}~\bibnamefont
  {Mukaiyama}}, \bibinfo {author} {\bibfnamefont {H.}~\bibnamefont {Katori}},
  \bibinfo {author} {\bibfnamefont {T.}~\bibnamefont {Ido}}, \bibinfo {author}
  {\bibfnamefont {Y.}~\bibnamefont {Li}}, \ and\ \bibinfo {author}
  {\bibfnamefont {M.}~\bibnamefont {Kuwata-Gonokami}},\ }\href {\doibase
  10.1103/PhysRevLett.90.113002} {\bibfield  {journal} {\bibinfo  {journal}
  {Phys. Rev. Lett.}\ }\textbf {\bibinfo {volume} {90}},\ \bibinfo {pages}
  {113002} (\bibinfo {year} {2003})}\BibitemShut {NoStop}%
\bibitem [{\citenamefont {Stellmer}\ \emph
  {et~al.}(2013{\natexlab{a}})\citenamefont {Stellmer}, \citenamefont {Grimm},\
  and\ \citenamefont {Schreck}}]{stellmer13}%
  \BibitemOpen
  \bibfield  {author} {\bibinfo {author} {\bibfnamefont {S.}~\bibnamefont
  {Stellmer}}, \bibinfo {author} {\bibfnamefont {R.}~\bibnamefont {Grimm}}, \
  and\ \bibinfo {author} {\bibfnamefont {F.}~\bibnamefont {Schreck}},\ }\href
  {\doibase 10.1103/PhysRevA.87.013611} {\bibfield  {journal} {\bibinfo
  {journal} {Phys. Rev. A}\ }\textbf {\bibinfo {volume} {87}},\ \bibinfo
  {pages} {013611} (\bibinfo {year} {2013}{\natexlab{a}})}\BibitemShut
  {NoStop}%
\bibitem [{\citenamefont {Loftus}\ \emph {et~al.}(2004)\citenamefont {Loftus},
  \citenamefont {Ido}, \citenamefont {Boyd}, \citenamefont {Ludlow},\ and\
  \citenamefont {Ye}}]{loftus04}%
  \BibitemOpen
  \bibfield  {author} {\bibinfo {author} {\bibfnamefont {T.~H.}\ \bibnamefont
  {Loftus}}, \bibinfo {author} {\bibfnamefont {T.}~\bibnamefont {Ido}},
  \bibinfo {author} {\bibfnamefont {M.~M.}\ \bibnamefont {Boyd}}, \bibinfo
  {author} {\bibfnamefont {A.~D.}\ \bibnamefont {Ludlow}}, \ and\ \bibinfo
  {author} {\bibfnamefont {J.}~\bibnamefont {Ye}},\ }\href {\doibase
  10.1103/PhysRevA.70.063413} {\bibfield  {journal} {\bibinfo  {journal} {Phys.
  Rev. A}\ }\textbf {\bibinfo {volume} {70}},\ \bibinfo {pages} {063413}
  (\bibinfo {year} {2004})}\BibitemShut {NoStop}%
\bibitem [{\citenamefont {Chaneli{\`e}re}\ \emph {et~al.}(2008)\citenamefont
  {Chaneli{\`e}re}, \citenamefont {He}, \citenamefont {Kaiser},\ and\
  \citenamefont {Wilkowski}}]{chaneliere08}%
  \BibitemOpen
  \bibfield  {author} {\bibinfo {author} {\bibfnamefont {T.}~\bibnamefont
  {Chaneli{\`e}re}}, \bibinfo {author} {\bibfnamefont {L.}~\bibnamefont {He}},
  \bibinfo {author} {\bibfnamefont {R.}~\bibnamefont {Kaiser}}, \ and\ \bibinfo
  {author} {\bibfnamefont {D.}~\bibnamefont {Wilkowski}},\ }\href {\doibase
  10.1140/epjd/e2007-00329-8} {\bibfield  {journal} {\bibinfo  {journal} {Eur.
  Phys. J. D}\ }\textbf {\bibinfo {volume} {46}},\ \bibinfo {pages} {507}
  (\bibinfo {year} {2008})}\BibitemShut {NoStop}%
\bibitem [{\citenamefont {Blatt}(2011)}]{blatt11}%
  \BibitemOpen
  \bibfield  {author} {\bibinfo {author} {\bibfnamefont {S.}~\bibnamefont
  {Blatt}},\ }\emph {\bibinfo {title} {Ultracold Collisions and Fundamental
  Physics with Strontium}},\ \href
  {https://jila.colorado.edu/publications/theses} {Ph.D. thesis},\ \bibinfo
  {school} {University of Colorado, Department of Physics} (\bibinfo {year}
  {2011})\BibitemShut {NoStop}%
\bibitem [{\citenamefont {Zelevinsky}\ \emph {et~al.}(2006)\citenamefont
  {Zelevinsky}, \citenamefont {Boyd}, \citenamefont {Ludlow}, \citenamefont
  {Ido}, \citenamefont {Ye}, \citenamefont {Ciury{\l}o}, \citenamefont
  {Naidon},\ and\ \citenamefont {Julienne}}]{zelevinsky06}%
  \BibitemOpen
  \bibfield  {author} {\bibinfo {author} {\bibfnamefont {T.}~\bibnamefont
  {Zelevinsky}}, \bibinfo {author} {\bibfnamefont {M.}~\bibnamefont {Boyd}},
  \bibinfo {author} {\bibfnamefont {A.}~\bibnamefont {Ludlow}}, \bibinfo
  {author} {\bibfnamefont {T.}~\bibnamefont {Ido}}, \bibinfo {author}
  {\bibfnamefont {J.}~\bibnamefont {Ye}}, \bibinfo {author} {\bibfnamefont
  {R.}~\bibnamefont {Ciury{\l}o}}, \bibinfo {author} {\bibfnamefont
  {P.}~\bibnamefont {Naidon}}, \ and\ \bibinfo {author} {\bibfnamefont
  {P.}~\bibnamefont {Julienne}},\ }\href {\doibase
  10.1103/PhysRevLett.96.203201} {\bibfield  {journal} {\bibinfo  {journal}
  {Phys. Rev. Lett.}\ }\textbf {\bibinfo {volume} {96}},\ \bibinfo {pages}
  {203201} (\bibinfo {year} {2006})}\BibitemShut {NoStop}%
\bibitem [{\citenamefont {Katori}\ \emph {et~al.}(1999)\citenamefont {Katori},
  \citenamefont {Ido}, \citenamefont {Isoya},\ and\ \citenamefont
  {Kuwata-Gonokami}}]{katori99}%
  \BibitemOpen
  \bibfield  {author} {\bibinfo {author} {\bibfnamefont {H.}~\bibnamefont
  {Katori}}, \bibinfo {author} {\bibfnamefont {T.}~\bibnamefont {Ido}},
  \bibinfo {author} {\bibfnamefont {Y.}~\bibnamefont {Isoya}}, \ and\ \bibinfo
  {author} {\bibfnamefont {M.}~\bibnamefont {Kuwata-Gonokami}},\ }\href
  {\doibase 10.1103/PhysRevLett.82.1116} {\bibfield  {journal} {\bibinfo
  {journal} {Phys. Rev. Lett.}\ }\textbf {\bibinfo {volume} {82}},\ \bibinfo
  {pages} {1116} (\bibinfo {year} {1999})}\BibitemShut {NoStop}%
\bibitem [{\citenamefont {Yang}\ \emph {et~al.}(2015)\citenamefont {Yang},
  \citenamefont {Pandey}, \citenamefont {Pramod}, \citenamefont {Leroux},
  \citenamefont {Kwong}, \citenamefont {Hajiyev}, \citenamefont {Chia},
  \citenamefont {Fang},\ and\ \citenamefont {Wilkowski}}]{yang15}%
  \BibitemOpen
  \bibfield  {author} {\bibinfo {author} {\bibfnamefont {T.}~\bibnamefont
  {Yang}}, \bibinfo {author} {\bibfnamefont {K.}~\bibnamefont {Pandey}},
  \bibinfo {author} {\bibfnamefont {M.~S.}\ \bibnamefont {Pramod}}, \bibinfo
  {author} {\bibfnamefont {F.}~\bibnamefont {Leroux}}, \bibinfo {author}
  {\bibfnamefont {C.~C.}\ \bibnamefont {Kwong}}, \bibinfo {author}
  {\bibfnamefont {E.}~\bibnamefont {Hajiyev}}, \bibinfo {author} {\bibfnamefont
  {Z.~Y.}\ \bibnamefont {Chia}}, \bibinfo {author} {\bibfnamefont
  {B.}~\bibnamefont {Fang}}, \ and\ \bibinfo {author} {\bibfnamefont
  {D.}~\bibnamefont {Wilkowski}},\ }\href@noop {} {\bibfield  {journal}
  {\bibinfo  {journal} {Eur. Phys. J. D}\ }\textbf {\bibinfo {volume} {69}}
  (\bibinfo {year} {2015})}\BibitemShut {NoStop}%
\bibitem [{\citenamefont {Stellmer}\ \emph
  {et~al.}(2013{\natexlab{b}})\citenamefont {Stellmer}, \citenamefont
  {Pasquiou}, \citenamefont {Grimm},\ and\ \citenamefont
  {Schreck}}]{stellmer13b}%
  \BibitemOpen
  \bibfield  {author} {\bibinfo {author} {\bibfnamefont {S.}~\bibnamefont
  {Stellmer}}, \bibinfo {author} {\bibfnamefont {B.}~\bibnamefont {Pasquiou}},
  \bibinfo {author} {\bibfnamefont {R.}~\bibnamefont {Grimm}}, \ and\ \bibinfo
  {author} {\bibfnamefont {F.}~\bibnamefont {Schreck}},\ }\href {\doibase
  10.1103/PhysRevLett.110.263003} {\bibfield  {journal} {\bibinfo  {journal}
  {Phys. Rev. Lett.}\ }\textbf {\bibinfo {volume} {110}},\ \bibinfo {pages}
  {263003} (\bibinfo {year} {2013}{\natexlab{b}})}\BibitemShut {NoStop}%
\bibitem [{\citenamefont {Nagel}\ \emph {et~al.}(2005)\citenamefont {Nagel},
  \citenamefont {Mickelson}, \citenamefont {Saenz}, \citenamefont {Martinez},
  \citenamefont {Chen}, \citenamefont {Killian}, \citenamefont {Pellegrini},\
  and\ \citenamefont {C\^ot\'e}}]{nagel05}%
  \BibitemOpen
  \bibfield  {author} {\bibinfo {author} {\bibfnamefont {S.~B.}\ \bibnamefont
  {Nagel}}, \bibinfo {author} {\bibfnamefont {P.~G.}\ \bibnamefont
  {Mickelson}}, \bibinfo {author} {\bibfnamefont {A.~D.}\ \bibnamefont
  {Saenz}}, \bibinfo {author} {\bibfnamefont {Y.~N.}\ \bibnamefont {Martinez}},
  \bibinfo {author} {\bibfnamefont {Y.~C.}\ \bibnamefont {Chen}}, \bibinfo
  {author} {\bibfnamefont {T.~C.}\ \bibnamefont {Killian}}, \bibinfo {author}
  {\bibfnamefont {P.}~\bibnamefont {Pellegrini}}, \ and\ \bibinfo {author}
  {\bibfnamefont {R.}~\bibnamefont {C\^ot\'e}},\ }\href {\doibase
  10.1103/PhysRevLett.94.083004} {\bibfield  {journal} {\bibinfo  {journal}
  {Phys. Rev. Lett.}\ }\textbf {\bibinfo {volume} {94}},\ \bibinfo {pages}
  {083004} (\bibinfo {year} {2005})}\BibitemShut {NoStop}%
\bibitem [{\citenamefont {Mickelson}\ \emph {et~al.}(2005)\citenamefont
  {Mickelson}, \citenamefont {Martinez}, \citenamefont {Saenz}, \citenamefont
  {Nagel}, \citenamefont {Chen}, \citenamefont {Killian}, \citenamefont
  {Pellegrini},\ and\ \citenamefont {C\^ot\'e}}]{mickelson05}%
  \BibitemOpen
  \bibfield  {author} {\bibinfo {author} {\bibfnamefont {P.~G.}\ \bibnamefont
  {Mickelson}}, \bibinfo {author} {\bibfnamefont {Y.~N.}\ \bibnamefont
  {Martinez}}, \bibinfo {author} {\bibfnamefont {A.~D.}\ \bibnamefont {Saenz}},
  \bibinfo {author} {\bibfnamefont {S.~B.}\ \bibnamefont {Nagel}}, \bibinfo
  {author} {\bibfnamefont {Y.~C.}\ \bibnamefont {Chen}}, \bibinfo {author}
  {\bibfnamefont {T.~C.}\ \bibnamefont {Killian}}, \bibinfo {author}
  {\bibfnamefont {P.}~\bibnamefont {Pellegrini}}, \ and\ \bibinfo {author}
  {\bibfnamefont {R.}~\bibnamefont {C\^ot\'e}},\ }\href {\doibase
  10.1103/PhysRevLett.95.223002} {\bibfield  {journal} {\bibinfo  {journal}
  {Phys. Rev. Lett.}\ }\textbf {\bibinfo {volume} {95}},\ \bibinfo {pages}
  {223002} (\bibinfo {year} {2005})}\BibitemShut {NoStop}%
\end{thebibliography}

%

\end{document}